# Co-doping of silicate bioceramics as a potential strategy to further enhance mono-doping consequences

Sohrab Mofakhami, Erfan Salahinejad*

Faculty of Materials Science and Engineering, K. N. Toosi University of Technology, Tehran, Iran

## Abstract

Silicate bioceramics have attracted significant attention in medical applications, particularly in hard tissue regeneration, because of their controllable chemical, physical, and biological functionalities, while ensuring biocompatibility. The coordination of silicate bioceramics with mono-dopants has been extensively studied to enhance their osteogenesis, angiogenesis, and antibacterial activity. However, the concept of employing dual or multiple co-doping to further enhance these biomaterials still demands more attention. This review paper originally focuses on the effect of chemical interactions among co-dopants and the principal constituents of the silicates, on the bio-behavior of these bioceramics. Additionally, future prospects of co-doped silicate bioceramics are outlined, including *in vivo* studies, clinical trials, and potential commercialization.

* Corresponding Author: Email Address: <salahinejad@kntu.ac.ir>

## Table of Content

# 1. Introduction

Bioceramics have extensive applications in biomedical fields, particularly in hard tissue reconstruction, therapeutic delivery systems, and polymer-matrix composite biomaterials. The bio-performance of bioceramics, including bioactivity, bioresorbability, biocompatibility, and antibacterial properties heavily depends on their composition, crystallinity, and specific surface area. As a distinct group of bioceramics, silicates are characterized by a network former of $SiO_2$ along with network modifiers like $CaO$, $Na_2O$, $P_2O_5$, and $MgO$. The adjacent building units of silicates are connected by bridging oxygens [1], where they are referred to as $Q_{Si}^n$ (n = 0-4) with "n" indicating the number of these atoms in each unit (Figure 1). However, the introduction of the network modifiers disrupts the lattice by substituting bridging oxygens with non-bridging ones, significantly impacting the properties of silicate bioceramics.

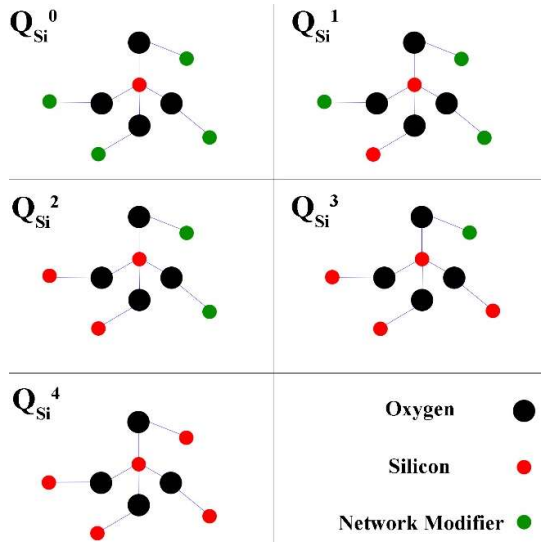

Figure 1. Constituting units of silicates.

The first silicate ceramic utilized in biomedical applications was 45S5 bioglass [2]. Nowadays, as well as bioactive silicate glass-ceramics and glasses with different compositions, crystalline silicates like calcium silicate (CaSi), wollastonite ($CaSiO_3$), forsterite ($Mg_2SiO_4$), akermanite ($Ca_2MgSi_2O_7$), bredigite ($Ca_7MgSi_4O_{16}$), monticellite ($CaMgSiO_4$), diopside ($CaMgSi_2O_6$), merwinite ($Ca_3MgSi_2O_8$), baghdadite ($Ca_3ZrSi_2O_9$), nagelschmidtite ($Ca_7Si_2P_2O_{16}$), gehlenite ($Ca_2SiAl_2O_7$), and hardystonite ($Ca_2ZnSi_2O_7$) are of interest to researchers. The common characteristic of silicate bioceramics is their ability to release silicon, a fundamental substance in the human body. Typically, this element stimulates the osteogenic differentiation and proliferation of cells for bone formation [3-5].

Despite exhibiting considerable bio-performance, standard silicate bioceramics with routine network modifiers still require further improvements in both biological and mechanical aspects for clinical applications. One successful method to enhance the behavior of silicates is their incorporation or doping with appropriate ions. Generally, dopants are elements, ions, or molecules intentionally introduced into a host material in small quantities, typically varying from a few parts per million (ppm) to a small percentage, to precisely modify a particular aspect of the material's structure and functionality in a controlled manner. The incorporation of dopants into silicate bioceramics has been reviewed in several papers. Typically, Schatkoski et al. [6] have reviewed the most commonly used dopants in bioceramics; Rabiee et al. [7] have focused on the impact of single ion substitution on the bio-performance of glasses; Palakurthy et al. [3] have studied Zn-incorporated bio-silicas; Devi et al. [4] have allocated a section of their review to single ionic doping of Mg silicate bioceramics; Cacciotti [8] have reviewed the impact of bivalent cationic doping on the performance of glasses; and Mohammadi and Sepantfar [9] have explored mono-doped silicates for coating applications.

Mono-doping alone may not meet all satisfactory requirements in certain cases, and the properties of mono-doped bioceramics can be further altered by co-incorporation with other ions. The co-doping strategy has been considered and reviewed in other fields, such as optoelectronics, electrocatalysis, semi-conductors, wastewater treatment, sensors, and energy devices [10-13]. Despite the publication of several research articles on co-doping of silicates, the literature is devoid of a review that provides a comprehensive analysis and understanding of the current knowledge in this specific field. Accordingly, this paper aims to review co-doping in silicates used in biomedical applications with a focus on the effect of chemical interactions among co-dopants and the principal constituents of the silicates on the bio-behavior of these bioceramics, being finalized with some potential co-doping pairs for the future research.

## 2. Doping elements used in biomedical co-doped silicates

Before discussing co-doping in biomedical silicates, it is critical to overview the impact of each element used in co-doped silicates on both the human body's function and the properties of silicate bioceramics (Table 1). Essentially, the use of dopants in silicate bioceramics aims to provide three specific bioproperties: antibacterial activity, osteogenesis, and angiogenesis. It should be noted that some other dopants, such as iron (Fe), have been successfully incorporated into silicate bioceramics but have been only explored in the context of mono-doping. Therefore, they are excluded from consideration in this list.

Table 1. Characteristics of the dopants used in co-doped silicate bioceramics

| Dopant | Advantages | Disadvantages | References |
|--------|-----------|---------------|-----------|
| **Aluminum** | Enhances mechanical and antibacterial behaviors | Delays apatite formation | [14, 15] |
| **Boron** | Improves angiogenesis | Reduces cell proliferation at high doses | [16, 17] |
| **Cobalt** | Enhances angiogenesis and antibacterial activity | Cytotoxicity at high doses | [18] |
| **Copper** | Enhances angiogenesis, antibacterial, and mechanical properties | Cytotoxicity at high doses | [19, 20] |
| **Fluorine** | Induces apatite formation and structural stability | Toxicity at high doses | [21, 22] |
| **Lithium** | Improves antibacterial, mechanical, and cell proliferation properties | Decreases apatite deposition | [23, 24] |
| **Magnesium** | Enhances cell activity, apatite formation, angiogenesis, mechanical strength, and stability | Often exhibits anomalous and complex effects | [25, 26] |
| **Manganese** | Promotes bone regeneration, cell activity, and antibacterial effects | Cytotoxicity at high levels | [27, 28] |
| **Rare Earth Elements** | Induces fluorescent and magnetic properties, supports osteogenesis and angiogenesis | Potential reduction of bioactivity, depending on the type and content of the dopants | [29, 30] |
| **Selenium** | Induces bone formation, antibacterial, and anticancer properties | Toxicity at high doses | [31, 32] |
| **Silver** | Induces antibacterial activity | Cytotoxicity at high doses | [33] |
| **Sodium** | Increases bioactivity and cytocompatibility | Promotes the cracking of polymers when high-Na glasses are used to reinforce polymers, via increasing water uptake and swelling | [34, 35] |
| **Strontium** | Improves bone regeneration and angiogenesis | Causes skeletal abnormalities at overexposure | [36, 37] |
| **Zinc** | Enhances cell proliferation, differentiation, anti-inflammatory effects, bone formation, and wound healing | Toxicity at high doses with delayed apatite formation | [38, 39] |
| **Zirconium** | Enhances structural stability, cell activity, mechanical, and antibacterial properties | Decreases bioactivity | [40-42] |

## 2.1. Aluminum

The presence of aluminum (Al) in the blood of healthy individuals is typically below 10 µg/l. Levels exceeding 100 µg/l are considered toxic, and even low-level exposure to Al can contribute to the onset or progression of Alzheimer's disease [43, 44]. The literature does not address the doping of biomaterials with Al extensively. However, its main advantages lie in enhancing mechanical and antibacterial properties. Mubina et al. [14] explored the incorporation of Al into bioglass with no significant influence on apatite formation and cytocompatibility. However, it provided the bioglass with antibacterial effects and significantly improved mechanical strength from 75 MPa to 105 MPa.

## 2.2. Boron

Boron (B) is recognized as an essential element for the body, particularly for wound healing and bone health with an accumulation of 0.9 ppm in bone [45]. One notable feature of B, when used as a dopant in silicates, is its ability to stimulate vascularization through the mitogenic effect on endothelial cells and enhancing the release of pro-angiogenic cytokines [46]. It has been reported that doping 58S bioglass with B enhances degradability and ion release rate, resulting in improved bioactivity and angiogenic properties. The latter has also been confirmed for CaSi [17] and 45S5 bioglass [47].

## 2.3. Cobalt

Cobalt (Co) is found in the body in quantities of 1-2 mg, primarily in the kidneys and liver. However, the main contribution of Co in the body is through B12 vitamin (Cobalamin) that is essential for red blood cell and DNA synthesis [48]. The incorporation of Co doping in silicate bioceramics has been shown to enhance osteogenesis, angiogenesis, and antibacterial properties. Co induces hypoxia-inducible factor 1 alpha by triggering a hypoxic cascade through the inhibition of the prolyl hydroxylase enzyme. This cascade leads to the secretion of growth factors required for vascularization [18]. Additionally, it has been noted that Co doping decreases the grain size and enhances the density of akermanite, resulting in increased microhardness and elastic modulus. Co doping also decreases the degradation rate while promoting apatite formation [49]. The incorporation of Co ions into borosilicate glasses, similar to akermanite, enhances stability, angiogenesis, and bone formation without causing cytotoxicity [18].

## 2.4. Copper

Copper (Cu) is a crucial component of some enzymes that play vital roles in several biological processes. However, there is a threshold for the amount of Cu in the body, as excessive amounts can cause oxidative damage leading to neurodegenerative changes. The Cu level in the adult human body is lower than 100 mg with accumulation in the kidneys, brain, liver, and heart [50-52]. The incorporation of Cu into biomaterials, including silicate bioceramics, has been extensively studied. Typically, bioglass [53, 54], wollastonite [55], akermanite [56], silicocarnotite [20], CaSi [57], diopside [19], and Mg-Ca silicate [58] have been evaluated after doping with Cu, resulting in improved antibacterial and angiogenic properties.

### 2.5. Fluorine

Fluorine (F) is a key element found in bone, dentin, and enamel minerals. When used as a dopant, F can enhance biostability by inhibiting the formation of cavities in the alveolar bone and providing better resistance against acidic environments. The addition of F increases the likelihood of fluorapatite formation instead of carbonated apatite, resulting in improved stability and mechanical properties. It has been pointed out that low concentrations of F can offer antibacterial effects and promote osteoblast activity. However, high concentrations of F can hinder osteoblast activity. The toxicity threshold for F in humans is reported to be 5 mg/kg of the body weight, with 15 mg/kg considered lethal [59-61]. F has been utilized as a dopant in silicate bioceramics, such as Mg-containing glasses [21], bioglass [62], bredigite [63, 64], and diopside [22, 65]. These studies mainly demonstrate an enhancement in stability and apatite formation with a decline in the medium pH. Moreover, to mitigate the toxic effects of high F concentrations, an optimal incorporation amount of 1-1.5 mol% in $SiO_2$-CaO-MgO glass has been reported to achieve the highest cytocompatibility and bioactivity [21].

### 2.6. Lithium

Lithium (Li) is primarily used in the medical field for treating manic depression and increasing the number of white blood cells. One of the key advantages of using Li as a dopant is its inhibitory effect on glycogen synthase kinase-3β, an enzyme that can impede cell growth and protective mechanisms. This effect of lithium shows promise for its antibacterial properties and increasing the proliferation of stem cells involved in bone reconstruction [66]. In this regard, the effect of Li doping in 45S5 bioglass-ceramics has been investigated [67], demonstrating that it leads to a reduction in melting, glass transition, and onset crystallization temperatures. These reductions ultimately contribute to improved glass stability and a decrease in apatite formation.

### 2.7. Magnesium

Magnesium (Mg) is a vital substance in the body with significant physiochemical contributions, especially to the brain, heart, and skeletal muscles. The estimated content of Mg in the human body is around 24 g, with 99% of it accumulating in intracellular spaces. This makes Mg the second most abundant element in these spaces after potassium (K). Moreover, almost 50-60% of the body's Mg is found in bones, with a crucial contribution to mineralization. Mg is involved in more than 600 enzymatic reactions and is known to improve anti-inflammatory properties, cell adhesion, and structural stability [68, 69].

Previous studies have reported the incorporation of Mg as a dopant into silicate bioceramics [26, 70]. The impact of Mg doping highly depends on factors such as sintering temperature, amount of Mg, composition, and crystallinity of silicate bioceramics. For instance, Mg incorporation into wollastonite has been proven to retard grain growth and to increase density by converting open pores to closed ones. This leads to improved structural stability, reduced degradability, and enhanced mechanical properties. Additionally, the incorporation of low amounts of Mg (below 7%) appears to promote new bone growth and blood vessel formation while decreasing cell viability [26]. In contrast, incorporating Mg into CaSi bioceramics results in a reduction of compressive strength due to the deterioration of sinterability and the formation of crystalline structures composed of $\beta$-CaSi, $MgSiO_3$, and $Ca_7MgSiO_4$ [25]. These contradictory behaviors following Mg doping make it a complicated species for doping applications.

### 2.8. Manganese

Manganese (Mn) is a trace element with a content of about 20 mg in the body, primarily accumulated in bone, brain, pancreas, liver, and blood. Mn constitutes metalloenzymes and serves as the initiator of various reactions, including hydrolases, kinases, and decarboxylases. Additionally, Mn plays significant roles in bone formation, wound healing, and the immune system. Insufficient Mn levels can lead to skeletal abnormalities during bone development, while overexposure to Mn can cause neurological disorders [71, 72].

The incorporation of Mn into silicates, such as bioactive glasses [73] and CaSi [25], has been assessed due to its potential to promote bone regeneration. The literature also illustrates that Mn doping can enhance cytocompatibility, angiogenesis ability, cell adherence, upregulation of osteogenic marker genes, and antibacterial properties. However, it has been realized that high levels of Mn have a detrimental effect on cell activities, indicating its toxicity, for instance, when exceeding 50 $\mu g/cm^2$ for MG63 cells [73].

### 2.9. Rare earth elements

The application of rare earth elements like cerium (Ce), gadolinium (Gd), ytterbium (Yb), thulium (Tm), and europium (Eu) as dopants in biomaterials has been studied due to their synergistic magnetic and optical properties. These doped bioceramics have potential applications in multimodal imaging for preventive medicine. The main concern in this context is the biocompatibility of these doped samples [74, 75]. The single incorporation of rare earth elements into silicate bioceramics has received limited attention. Zambanini et al. [76] demonstrated that Gd or Yb doping into silicate bioceramics induce no detectable cytotoxicity, with the samples exhibiting over 80% cell viability. Furthermore, Borges et al. [77] showed that the incorporation of Gd or Yb can lead to increased non-bridged oxygen bands and reduced network connectivity, giving rise in a higher degradation rate and bioactivity. Shi et al. [29] also indicated that Eu-doped mesoporous silica nanospheres, in addition to their fluorescent property, induce osteogenesis and angiogenesis.

### 2.10. Selenium

Selenium (Se) naturally exists in the body at levels of 10-15 mg, mainly in selenoproteins such as glutathione peroxidase and deiodinases. It plays a critical role in various bodily functions, including antioxidant defense, proliferation of activated T cells, cancer prevention, DNA synthesis, and the formation of thyroid hormones [78]. Selenium is predominantly accumulated in the kidneys and liver, but it is found in all cells of the human body. Despite the aforementioned functions, the main appeal of selenium as a dopant is its antibacterial property and ability to promote bone formation [79, 80]. However, the investigation of Se-doped silicate bioceramics has been limited. Wang et al. [31] demonstrated that Se-doped mesoporous bioglasses exhibit a great apatite formation ability, along with a controlled release of doxorubicin. Similarly, Wu et al. [32] showed that Se-doped mesoporous bioglass nanospheres can promote apatite formation and controlled release of drugs. They also found that the doped samples have different cytotoxicity impacts on cancer and normal cells.

### 2.11. Silver

Silver (Ag) is not considered essential for the human body, while its trace amounts may be present in the body due to dietary intake. Ag has a detrimental impact on cells when released beyond the permissible limit, as it interferes with the activities of membrane proteins, thereby inhibiting cell growth [81, 82]. Ag interact with DNA, specifically inhibiting the DNA replication of bacteria and causing their death. Additionally, the reactivity of Ag ions results in the rupture of bacterial walls and the death of bacterial cells [83]. The incorporation of Ag into biomaterials, particularly silicate bioceramics [84], has been extensively evaluated due to its considerable effect on antibacterial properties, as well as its anti-inflammatory ability. It has been concluded that even low amounts of Ag, below the cytotoxicity threshold, can either maintain or enhance the apatite-forming ability and dramatically increase the antibacterial efficacy of silica bioceramics.

### 2.12. Sodium

The total content of sodium (Na) in a healthy adult human body is approximately 4200 mmol, predominantly distributed in the extracellular fluid and bone. Sodium is one of the most crucial substances for the human body, and a daily intake is essential for maintaining health. However, excessive sodium intake can lead to serious health conditions [85]. The presence of Na in the human body's bone suggests its potential to be used as a dopant in bioceramics to enhance osteogenesis properties. Typically, Rahmani and Salahinejad. [34] studied the use of sodium as a single dopant in diopside, indicating an enhancement in bioactivity and cytocompatibility due to this addition.

### 2.13. Strontium

Strontium (Sr) is of the most abundant substances found in certain tissues, particularly in bones. Nearly 99% of the body's Sr is accumulated in teeth and bones. While Sr provides advantages and is necessary for the body, excessive exposure to it has been shown to cause skeletal abnormalities in children [86]. Sr is commonly used as a dopant due to its significant potential for improving bone structure and inhibiting bone resorption. As a result, researchers have extensively investigated the incorporation of Sr into silicates, including bioglasses [87], baghdadite [88], forsterite [36], diopside [89], gehlanite [90], CaSi [91, 92], and tricalcium silicate [93]. Thus, Sr is perhaps the most popular element utilized for doping silicate bioceramics.

Regardless of the silicate composition, all Sr-doped silicate bioceramics exhibit bioactivity and biocompatibility without any cytotoxic effects. Moreover, Sr doping promotes the formation of apatite, cell activities, and alkaline phosphatase (ALP) activity, improving osteogenesis and angiogenesis. The bioresorbability of bioceramics depends on their form and composition. For instance, Sr incorporation reduces the degradability of baghdadite [88] and CaSi [91] scaffolds, while it increases the degradability of forsterite and CaSi cements [92]. In case of gehlanite, the total degradability shows no significant difference, and the only variation is associated with the amount of released ions [90]. The influence of Sr incorporation on mechanical properties also relies on the composition of silicate bioceramics. Studies have confirmed that Sr-doped gehlanite [90] exhibits inferior mechanical behaviors compared to non-doped samples, whereas the incorporation of Sr into CaSi [91] and baghdadite [88] improves mechanical properties.

### 2.14. Zinc

Zinc (Zn) is considered a vital substance in the body as it critically contributes to the function of over 50 metalloenzymes that affect cellular behavior. The adult human body consists of almost 1.5-2.5 g of zinc, mostly found intracellularly, particularly in bone (86%), skin (4.2%), liver (3.4%), and red blood cells. Zinc deficiency in the body can give rise to a decrease in bone weight and bone formation [94, 95]. However, the most notable attribute of Zn is its involvement in gene expression and cell proliferation. Additionally, Zn possesses anti-inflammatory, antibacterial, bone-stimulating, and bone resorption-inhibiting properties, making it a widely investigated dopant in bioceramics.

Zn-doped silicate bioceramics, including bioglasses [39], forsterite [96], and tricalcium silicate [38] have been studied in medical applications. These doped ceramics demonstrate proper cytocompatibility, antibacterial ability, osteoblast differentiation, and bone regeneration. However, the release of high levels of Zn can be cytotoxic, for example, 2.7 ppm for endothelial cells [97]. Also, excessive Zn amounts can negatively affect bioactivity at levels exceeding 10 mol% [8]. The impact of Zn doping on degradability and apatite formation exhibits contrasting behaviors. In most cases, such as akermanite, bioglasses, and CaSi, the incorporation of Zn leads to a decline in the degradation rate owing to the high affinity of Zn with the constituent ions of the silicates [39]. However, Zn doping enhances the degradation rate of forsterite, which is attributed to reduced grain size and increased porosity [96].

### 2.15. Zirconium

The body consists of almost 300 mg of zirconium (Zr), with the main accumulation found in the heart, kidney, liver, and lungs. Zr has been utilized in various applications, including hemodialysis, peritoneal dialysis, deodorant preparation, production of wearable synthetic kidneys, and hip replacements. Zr has been proven to be biocompatible with low toxicity [98]. The incorporation of Zr into bioglasses has been investigated [42], indicating that the doped bioglasses have superior stability and mechanical properties compared to non-doped counterparts. They also promote apatite formation and cell proliferation, differentiation, and antibacterial properties.

## 3. Dual co-dopants used in biomedical silicates

Even with a comprehensive understanding of the individual effects of dopants, there is still a need for theoretical, computational, and/or experimental investigations to characterize co-doped bioceramics due to interactions between dopants. This section provides a brief review of the literature on co-doped bioceramics, as listed in Table 2.

Table 2. A review on the influence of dual co-doping on the biological characteristics of silicates

| Co-dopants | Host | Advantages | Disadvantages | References |
|---|---|---|---|---|
| **Al-Mg** | 45S5 bioglass | Increases mechanical properties | Decreases bioactivity | [99] |
| **Al-Sr** | 45S5 bioglass | - | Lowers antibacterial and apatite formation abilities | [100] |
| **B-Co** | 45S5 bioglass | Induces apatite formation and angiogenesis ability | - | [46] |
| **B-Cu** | 62% $SiO_2$-9% $P_2O_5$-9% CaO-5% CuO-15% $B_2O_3$ | Promotes apatite formation | - | [101, 102] |
| **B-Mg** | CaSi | Increases apatite formation and compression strength | - | [103] |
| **B-Na** | 13.27% MgO–30.97% CaO–2.65% $P_2O_5$–39.82% $SiO_2$–4.43% $B_2O_3$–4.43% $Na_2O$–4.43% $CaF_2$ | Improves apatite formation | - | [35] |
| **Ce-Ag** | 80S bioglass | Endows antibacterial effect | Cytotoxicity at high dosages of Ag and reduced apatite formation ability at high dosages of Ce | [104] |
| **Co-Li** | 70% $SiO_2$–4% $P_2O_5$–26% CaO | Enhances antibacterial effect, cell activity, osteogenesis, and angiogenesis | Cell viability and proliferation lower than Li mono-doping | [105] |
| **Co-Sr** | 41.20% $SiO_2$–5.06% $P_2O_5$–36.14% CaO–7.17% $Na_2O$–3.26% MgO–7.17% $K_2O$ glass | Enhances osteogenesis, angiogenesis, and collagen fiber formation | Decreases cell activity, osteogenesis, and angiogenesis | [106-108] |
| | 58% $SiO_2$–26% CaO–9% $P_2O_5$–5% SrO–2% CoO | Enhances cell attachment, proliferation, and osteogenic gene expression | - | [109] |
| **Cu-Ag** | 82S5 bioglass | Provides apatite formation and antifungal property | - | [110] |
| **Cu-Sr** | Diopside | Enhances mechanical properties, cell viability, and antibacterial property | - | [111] |
| | 80% $SiO_2$–(15-2x)% CaO–x% CuO–x% SrO–5% $P_2O_5$ | Increases specific surface area and antibacterial property | Retards apatite formation and lower cell viability | [112] |
| | 50S6P | Improves apatite formation ability | Delays apatite formation due to the presence of Cu | [113] |
| | $\beta$-$Ca_{1.5}$ $Sr_{0.3}$ $Cu_{0.2}$ $SiO_4$ | Improves osteogenesis and angiogenesis ability | - | [114] |
| | 58% $SiO_2$–26% CaO–9% $P_2O_5$–5% SrO–2% CuO | Enhances cell attachment, proliferation, and osteogenic gene expression | - | [109] |
| **Er-Yb** | $CaSiO_3$ glass | Induces luminescence property | - | [75] |
| | Glassy akermanite | Induces luminescence property | Declines apatite formation | [74] |
| | 70% $CaSiO_3$–30% $TiO_2$ glass | Induces luminescence property | Declines apatite formation | [115] |
| **Eu-Gd** | 60% $SiO_2$–4% $P_2O_5$–28% CaO–8% Eu glass | Induces fluorescence and biocompatibility | - | [116] |
| **Eu-Sr** | $\beta$-$Ca_{1.7}$ $Eu_{0.1}$ $Sr_{0.2}$ $SiO_4$ | Enhances osteogenesis, angiogenesis, compression strength, and drug release behavior | - | [114] |
| **F-Sr** | Diopside | - | Decreases apatite formation and cell activity | [117] |
| **Li-Ag** | 58S bioglass | Enhances apatite formation, cell activity, and antibacterial properties | - | [118] |
| | 68S bioglass | Induces ALP activity and antibacterial property | - | [119] |
| | 58S and 68S bioglass | Increases cell activity and antibacterial property | Cytotoxicity at high levels of Ag with a threshold of 10% | [120] |
| **Mg-Ag** | 35.5% $SiO_2$–4% $P_2O_5$–40% CaO–20.5% $Na_2O$ | Enhances apatite formation and antibacterial property | Deteriorates cell proliferation | [121] |
| | 60% $SiO_2$–31% CaO-4% $P_2O_5$–5% SrO | Induces antibacterial properties | Delays apatite formation and cytotoxicity at high amounts of Ag | [122] |
| **Mg-Sr** | Wollastonite | Enhances degradability, apatite formation, and cell activity | Decreases mechanical properties | [123] |
| | Hardystonite | High cell viability and proliferation | - | [124] |
| | 45S5 bioglass | Enhances apatite formation, cytocompatibility, cell proliferation, and ALP activity | - | [125] |
| | Wollastonite | Promotes angiogenesis and osteogenesis | - | [126] |
| **Mg-Zn** | $Ca_2SiO_4$ | Enhances apatite formation and cell viability | Deteriorates cell viability at high levels | [127] |
| | 71% $SiO_2$–26% CaO–5% $P_2O_5$ glass | - | Decreases apatite formation | [128] |
| | 46.1% $SiO_2$–26.9% CaO–2% MgO–2.6% $P_2O_5$–(22.4-x)% $Na_2O$-x% ZnO | - | Reduces apatite formation by increasing the Zn value | [129] |
| **Mn-Zn** | 43% CaO–42% $SiO_2$–9% $P_2O_5$–6% ZnO glass | Enhances cell viability and degradability | - | [130] |
| **Se-Sr** | 43% CaO–40% $SiO_2$–12% $P_2O_5$–5% MgO glass | Enhances degradability, apatite formation, cell activity, antibacterial, and anticancer properties | - | [131, 132] |
| **Ag-Sr** | S53P4 Bioglass | Enhances apatite formation, cell viability, and proliferation | Decreases antibacterial property | [133] |
| | 70% $SiO_2$–30% CaO | Enhances apatite formation and antibacterial properties | - | [83] |
| | $\beta$-$Ca_{1.85}$ $Sr_{0.1}$ $Ag_{0.05}$ $SiO_4$ | Enhances osteogenesis and angiogenesis ability | - | [114] |
| **Ag-Zn** | 78S bioglass | Enhances antibacterial effect and cytocompatibility | Lowers apatite formation ability compared to non-doped and Ag mono-doped samples | [134] |
|  | $Sr_5(PO_4)_2SiO_4$ | Improves antibacterial effect against E. coli bacteria, hemocompatibility, and apatite formation ability | Cytotoxicity and non-hemocompatibility at high levels of doping, | [135] |
|  | 46.48% $SiO_2$–17.5% CaO–35.9% SrO | Antibacterial effect, cell differentiation, cytocompatibility till a certain threshold depending on type of cell, and ability to kill bone cancer cells | Decreases Cytocompatibility and mineralization at high concentrations of Ag | [136] |
| **Ag-Zr** | 58S bioglass | Improves apatite formation, cell activity, and antibacterial property | - | [137] |
| **Sr-Zn** | Akermanite | Enhances cytocompatibility and ALP activity | - | [138] |
|  | 49.46% $SiO_2$–27.27% CaO–6.60% $Na_2O$–6.60% $K_2O$–1.07% $P_2O_5$–3% MgO–3% SrO–3% ZnO nanofibers | Considerable cytocompatibility and antibacterial property | - | [139] |
|  | (36.07–x)% CaO–x% SrO–(19.24-x)% MgO–x% ZnO 5.61% $P_2O_5$ 38.49% $SiO_2$–0.59% $CaF_2$ | Enhances cell viability and proliferation up to a certain level of dopants, antibacterial property | Decreases apatite formation | [140] |
| | $\beta$-$Ca_{1.6}$ $Sr_{0.3}$ $Zn_{0.1}$ $SiO_4$ | Improves osteogenesis and angiogenesis ability | - | [114] |
| | Wollastonite | Improves cytocompatibility, osteogenesis, and angiogenesis ability | Reduces apatite formation | [141] |
| **Zr-Zn** | 60% $SiO_2$-5% $ZrO_2$-4% $P_2O_5$-(31-x)% CaO-x% ZnO | Promotes cell viability, proliferation, and antibacterial effect | Reduces the safe amount of Zn after Zr addition | [142] |

### 3.1. Aluminum-Magnesium

Al-Mg co-doping has been investigated in non-medical applications, such as photovoltaic energy harvesting devices [143], ZnO gas sensors [144], photoconductive ZnO films [145], and lithium ion batteries [146]. However, there is a lack of reports on the use of Al-Mg doping in non-silicate biomaterials. Regarding silicates, Karakuzu-Ikizler et al. [99] focused on fabricating Al-Mg co-doped 45S5 bioglass, where each dopant was present at a concentration of 1 mol%. In this study, Mg was substituted with Ca, and Al with Na in the glass structure. The co-doped sample exhibited the highest Vickers hardness both prior to and following immersion in the simulated body fluid (SBF), compared to the samples with Mg mono-doping and non-doped samples. After 7 days of incubation, apatite-like crystals were precipitated on all of the samples, albeit in different shapes and sizes. Interestingly, the non-doped samples showed the highest bioactivity, while the Mg mono-doped samples showed the least bioactivity.

Based on the available information, it is challenging to definitively determine whether Al-Mg co-doping into silicates is favorable for biomedical applications, as research on this specific pair is still limited. However, considering the fact that non-doped samples exhibited the highest bioactivity in the aforementioned work, it is inferred that the use of Al-Mg co-doping may not be logical unless the improvement of mechanical properties is a primary concern, particularly for load-bearing purposes. Nonetheless, further focuses are needed to fully understand the benefits and drawbacks of Al-Mg co-doping in silicate-based biomaterials.

### 3.2. Aluminum-Strontium

There are no reports on the combination of Al and Sr as dopants in either non-biomedical or biomedical applications, except in case of silicate bioceramics, to our knowledge. Araujo et al. [100] focused on the impact of Al-Sr co-doping into 45S5 bioglass. It was detected that following 7 days of incubation in SBF, all the specimens, except the Al mono-doped sample, effectively stimulated apatite formation. However, the deposition of apatite on the Al mono-doped sample after 14 days indicated that Al does not completely inhibit but rather delays the apatite formation process. The antibacterial influence against *E. coli* bacteria was observed for all the samples, but the Al-doped and Al-Sr co-doped samples required higher doping concentrations (4-8 ppm) to reach a minimum inhibitory concentration compared to the non-doped and Sr mono-doped samples (2-4 ppm). This can be attributed to the retarding effect of Al on the degradation of bioceramics. The NCTC clone 929 cell viability of all the samples was above 90%, indicating the cytocompatibility of the samples.

Despite the cytocompatibility of Al-Sr co-doped 45S5 bioglass, the Al-Sr combination cannot outperform many other doping pairs due to the slow degradation of Al-containing samples, which can be problematic in clinical applications. Moreover, the existing results on this pair are insufficient to confidently assess its capabilities, necessitating further experiments. Therefore, until more results are obtained, and considering that there are other doping pairs that exhibit similar attributes with superior performance, Al-Sr cannot be considered a highly promising co-doping strategy for future biomedical applications, particularly in the context of bone regeneration.

### 3.3. Boron-Cobalt

Previous studies have utilized B and Co co-doping to enhance the durability of $TiO_2$ anodes in water treatment applications [147]. In biomedicine, a scaffold for tissue regeneration should not only promote tissue regrowth but also enhance angiogenesis. New blood vessels provide necessary nutrients and oxygen to the site, facilitate the circulation of progenitor cells, and aid in the removal of metabolic waste. Considering the angiogenesis-stimulating properties of B and Co, researchers have explored their incorporation into silicates from this perspective.

Chen et al. [46] produced B-Co co-doped 45S5 bioglass through the melt-quenching process to evaluate bioactivity and angiogenesis. The co-doped samples demonstrated the ability to induce the bioactive precipitation of hydroxycarbonate apatite (HCA). While Co delayed the formation of HCA, B enhanced it. The release behavior of the co-doped specimens showed that higher B levels resulted in a greater release of Si ions. This is owing to the introduction of B into the silica network, which reduces its connectivity and stability. As a result, P and Ca ions remaining in the medium decreased due to the improved CHA precipitation, indicating an increased release of Si. On the other hand, an increase in the Co content led to higher levels of P and Ca ions in the medium, suggesting the inhibitory effect of Co on the CHA formation due to reduced degradability. This behavior was confirmed by the study of Co mono-doped akermanite [49]. Cytocompatibility was found to depend strongly on the amount of these ions in the medium. The media containing 0.1 or 1% extracts of the different samples showed no cytotoxicity on MG-63 cells and displayed morphologies comparable to those of the positive control. However, the media containing 10% extract of all the samples exhibited cytotoxicity (Figure 2a). The threshold level of B and Co for cytotoxicity typically ranges from 0.65 mM for MLO-A5 cells to 100 µM for MG-63 cells [148, 149]. Additionally, the co-doped glass showed an improved secretion of vascular endothelial growth factor (VEGF) compared to the non-doped and mono-doped counterparts. This enhanced release behavior can be owing to the synergistic influence of the dopants, especially in the culture medium containing 1% extraction of the samples in comparison to 0.1% (Figure 2b).

**(a)**

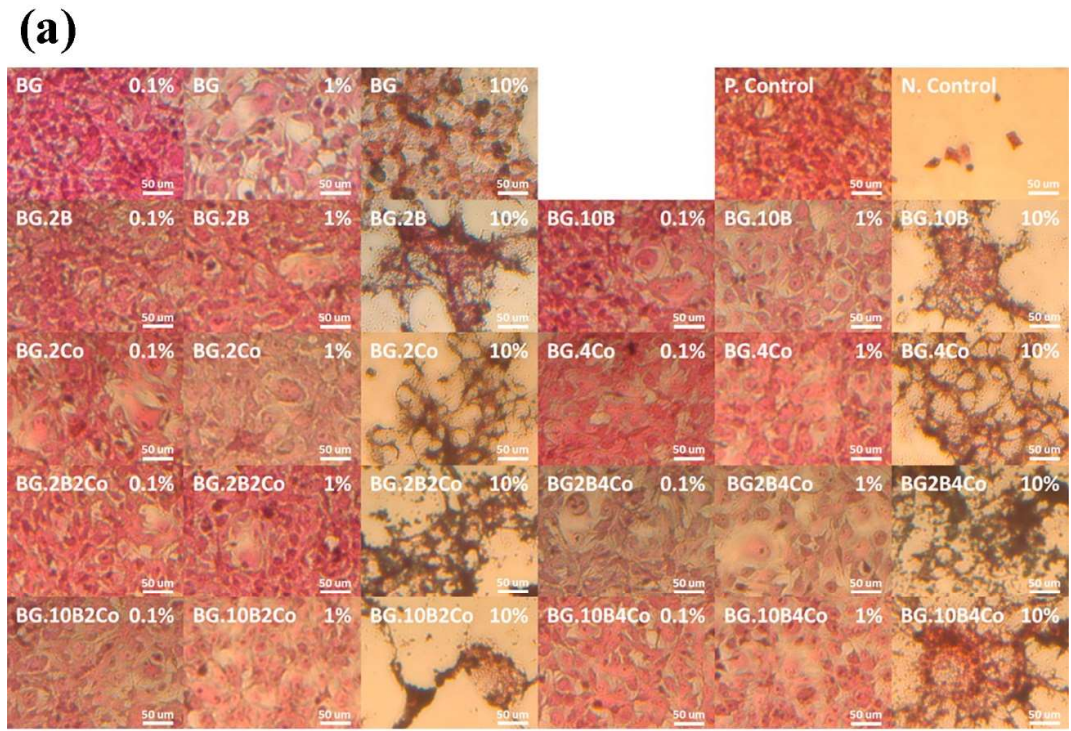

**(b)**

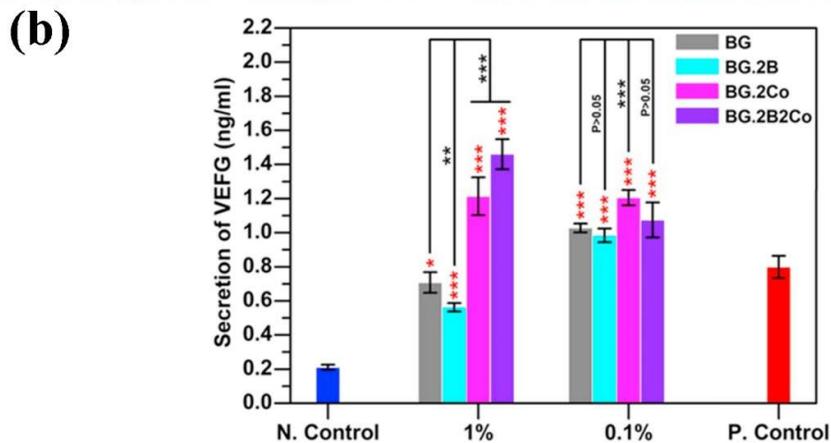

Figure 2. Images of MG-63 cells cultured in differently concentrated extractions of non-, B-, Co-, co-doped 45S5 bioglasses, negative and positive control samples (a). VEGF secretion from stromal cells cultured with 0.1 and 1% extractions of the non-, mono-, and B-Co co-doped 45S5 bioglasses (b), reprinted with permission from Ref. [46].

In conclusion, while B-Co doping has a positive impact on vascularization, it is essential to carefully control the concentrations of the dopants to ensure biocompatibility and bioactivity. The B-Co pair holds significant potential for biomedical applications, particularly in bone reconstruction. The combined effect of B and Co on angiogenesis can effectively enhance the delivery of nutrients and oxygen. Furthermore, the antibacterial properties of this dopant pair can facilitate a smoother healing process and potentially reduce the need for administering drugs to prevent infection. However, it is important to emphasize that realizing these benefits requires the rigorous execution of comprehensive experiments, particularly under *in vivo* conditions.

### 3.4. Boron-Copper

The use of B and Cu as co-dopants in non-biomedical applications has been examined by Wang et al. [150] focusing on the magnetocaloric properties of $Mn_{1.05} Fe_{0.9} P_{0.5-x} Si_{0.5} Cu_{0.10}$. Similarly, Cao et al. [151] investigated the impact of this co-doping on the photocatalytic properties of ZnO. In biomedical applications, B and Cu are both believed to enhance the vascularity of bioceramics when used as dopants. Moreover, Cu has been proven to have an antibacterial effect. However, to the best of our knowledge, no research has been conducted on the effect of this pair on the biomedical properties of ceramics other than silicate bioceramics. Typically, Miola et al. [101] characterized B-Cu co-doped 62% $SiO_2$–9% $P_2O_5$–9% CaO-5%

CuO-15% $B_2O_3$ bioglass, showing the improved apatite formation ability of the co-doped sample compared to the non-doped sample. This superior behavior is believed to be due to the effect of B that can increase bioactivity. These results were also confirmed by Piatti et al. [102].

In conclusion, B-Cu co-doping at optimal levels can improve bioactivity, making it suitable for bone applications. However, further tests, such as antibacterial activity, cellular activity, angiogenesis, and *in vivo* trials are needed to further clarify the abilities of this pair.

### 3.5. Boron-Magnesium

The employment of B-Mg co-doping in non-biomedical applications has been addressed to increase the stability of nickel-rich cathode materials [152] and to improve the thermochromic behaviors of $BiVO_4$ [153]. However, as of our knowledge, there are no studies on the biomedical applications of this pair, excluding silicate bioceramics. Ru [103] developed CaSi 3D-printed porous constructs doped with B and Mg at 0.78 and 1.18%, respectively. The co-doped samples exhibited a compression strength of 50 MPa, which was higher than that of non-doped sample. In addition, after soaking the samples in SBF for 60 and 168 h, a biological osteoid apatite layer was observed on their surface, indicating the bioactivity of these co-doped powders.

Due to the lack of enough information, it is not possible to reach a firm conclusion on how this pair will act as a co-doping option. To understand if B and Mg can be considered for development, there is still a long way and many things should be cleared such as at how they interact with cells and other biological factors, their angiogenesis and osteogenesis ability, and their optimized value for cytotoxicity as B can be highly toxic at high doses. Therefore, this pair is not yet to be considered as a potential pair for biomedical applications nor is ruled out.

### 3.6. Boron-Sodium

In non-biomedical applications, graphitic carbon nitride co-doped with B and Na has been produced to alter the photocatalytic activity of $H_2O_2$ production [154], photocatalytic hydrogen generation in carbon nitride nanosheets [155], and the stability and variation rate of ZnO varistors [156]. However, to our knowledge, there are no reports of using this pair as a co-dopant in non-silicate bioceramics. In the context of silicate bioceramics, Da Fonte and Goel [35] produced a B-Na co-doped bioactive glass with the composition of 13.27% MgO-30.97% CaO-2.65% $P_2O_5$-39.82% $SiO_2$-4.43% $B_2O_3$-4.43% $Na_2O$-4.43% $CaF_2$, showcasing the formation of HCA layer after 24h of immersion in SBF.

Currently, the ability of B-Na co-doping to induce biomineralization *in vitro* is the only available knowledge on silicate bioceramics. Given this limited information, it is difficult to predict the future of this pair in biomedical applications. There are many more factors that need to be evaluated before proceeding with this pair. For instance, the cytotoxicity effect should be considered as B can be harmful to certain cells at high doses. Additionally, the angiogenesis property of this pair needs to be investigated.

### 3.7. Cerium-Silver

The use of Ce-Ag co-doping in non-biomedical applications has been studied to modify the photocatalytic effects of ZnO [157] and $TiO_2$ [158]. However, to our knowledge, their application in non-silicate biomaterials has not yet been reported. Taye et al. [104] investigated the effect of Ce-Ag co-doping on the properties of 80S bioglasses. The co-doped sample contained 2.87 mol% of each dopant, while four mono-doped samples were prepared with either 2.87 or 5.75% of each dopant. The antibacterial effect of these doped samples was tested against *E. coli* bacteria, resulting in 100% bacteria killed after 24 h for the co-doped sample and the 5.75% Ag mono-doped sample. The 2.87% Ag mono-doped sample was in the third place, killing 86.6% of *E. coli* bacteria due to the more effectiveness of Ag compared to Ce. However, co-doping with Ce had the same effect as doubling the amount of Ag, which is advantageous considering that high dosages of Ag can be cytotoxic. After immersion of the samples in SBF for seven days, all the samples, except for the 5.75% Ce mono-doped glasses, were able to promote the development of HA crystals on their surface. This indicates that high values of Ce can leave a negative effect on bioactivity. MC3T3-E1 osteoblasts viability tests indicated that all the samples, except for the 5.75% Ag-doped sample (about 70% cell viability) were cytocompatible. The highest cell viability was for the 5.75% Ce mono-doped with about 100% cell viability. The cell viability of the co-doped sample was measured to be almost 87%.

In conclusion, the available results are still insufficient for a definitive decision, as there is no real data on the bone regeneration capability of these co-doped samples or enough data on how they interact with different types of silicate bioceramics. More importantly, no *in vivo* data has been provided until now. However, based on the current results, it can be said that this pair has an additive effect, partially compensating for each other's disadvantages. For instance, high values of Ag can lead to cell death, whereas for the most effective antibacterial results, high dosages of Ag are required. However, the addition of Ce could boost the antibacterial effect of Ag, making it possible to achieve the same results with less Ag, while simultaneously increasing the cell activity. Therefore, this pair, both employed at low dosages, could be valuable for future research and bone regenerative applications, especially when dealing with small bone defects.

### 3.8. Cobalt-Lithium

In non-biomedical applications, the effect of co-doping with Co and Li has been investigated to enhance the performance of nickel oxide solar cells [159]. In biomedical applications, Co-Li co-doping into ZnO has been investigated for altering optical properties and providing antibacterial effects [160]. Regarding silicates, Zhang et al. [105] investigated the effect of this co-doping into bioglass nanoparticles with the nominal composition of 70% $SiO_2$-4% $P_2O_5$-(26-X-Y) CaO-X% $Li_2O$-Y% CoO where X= 0 or 5, and Y= 0 or 2. The apatite formation ability in SBF after seven days indicated that the Li-doped and non-doped samples were able to from a cauliflower-like structure, whereas the Co-doped and co-doped samples had needle-like crystals. This needle-like structure was observed on the Li mono-doped and non-doped samples only after 3 days, indicating the faster ability of these two samples to form an apatite layer. *E. coli* and *S. aureus* bacteria were able to grow well in the presence of the non-doped and Li doped samples, whereas in case of Co-doped and co-doped samples, their growth was significantly reduced. This suggests the effect of Co as an antibacterial dopant and the non-effectiveness of Li in the bacterial resistivity. Proliferation and viability for bone marrow stromal cells (BMSCs) after one day were independent of the concentration of bioglass extracts. However, after five days or for higher concentrations, the Li-doped samples had the highest proliferation, whereas the Co-doped sample had the lowest. Moreover, the difference between the co-doped and non-doped samples were insignificant and slightly lower than the Li-doped sample and higher than the Co-doped sample. This indicates that the addition of Li was able to partially compensate for the cytotoxicity of Co. The samples containing Li were more effective in terms of osteogenesis, while the samples containing Co were more effective in angiogenesis properties, with the co-doped samples showing slightly the best performance in both cases. These results indicate that the additive effect of these dopants in terms of vascularization and bone formation ability can make them a great candidate for bone regenerative applications.

In conclusion, the results from the work of Zhang et al. [105] seem to promise an effective pair, especially in case of bone regenerative applications. *In vitro* osteogenesis and angiogenesis together can be an indicator of great ability for bone regeneration ability *in vivo*. Furthermore, antibacterial properties can further ease the repair process for patients. However, the *in vitro* results are far from sufficient for a firm decision on whether Co and Li can find their way to commercialization or not as *in vivo* tests are required.

### 3.9. Cobalt-Strontium

Co and Sr co-doping has been extensively employed in lanthanum chromite fuel cells [161]. In the field of biomedicine, Co and Sr have positive effects on angiogenesis and osteogenesis, respectively. This suggests that combining these elements through co-doping could lead to superior bio-performance. Consequently, researchers have been exploring the incorporation of Co and Sr into bioactive glasses.

Kermani et al. [106] focused on the biocompatibility and bone reconstruction capabilities of a mesoporous Co-Sr co-doped glass (41.20% $SiO_2$–5.06% $P_2O_5$–36.14% CaO–7.17% $Na_2O$–3.26% MgO–7.17% $K_2O$). While Co doping resulted in a slight decline in cell viability, co-incorporation with Sr improved cell cytocompatibility by neutralizing the adverse effects of Co. The non-doped glass exhibited the highest osteogenesis and angiogenesis abilities due to its higher Ca content compared to the doped specimens. Among the doped specimens, the Sr-doped glass showed superior angiogenesis ability. Similar conclusions were drawn by Kargozar et al. [107] on Co-Sr co-doped bioactive glasses (41.20% $SiO_2$–5.06% $P_2O_5$–36.14% CaO–7.17% $Na_2O$–3.26% MgO–7.17% $K_2O$), where Sr doping outperformed the other doped and non-doped states from the viewpoints of cell attachment, viability, proliferation, ALP activity, and VEGF release.

In another work done by Kargozar et al. [108], Co-Sr co-doping of 41.20% $SiO_2$–5.06% $P_2O_5$–36.14% CaO–7.17% $Na_2O$–3.26% MgO–7.17% $K_2O$ glasses was studied using human umbilical cord perivascular cells (HUCPVCs). The results depicted that the Sr-doped samples exhibited the best cell proliferation outcomes, attributed to the release of Co from the other samples. The expression of genes related to osteogenesis also indicated the superior performance of the co-doped specimen compared to the other specimens. In terms of VEGF expression, the Co-containing samples demonstrated better results compared to the other specimens, whereas there was no meaningful difference between the Co and Co-Sr doped samples. *In vivo* tests were also conducted using a large bone defect rabbit model. After 4 weeks of implantation, bone formation occurred on all of the specimens, with no evidence of chronic inflammation. The remaining bioactive glass was identifiable in all the samples after 4 and 12 weeks, but the reduction in the particle size after 12 weeks indicated the biodegradability of the glass samples. Notably, the co-doped samples exhibited a higher level of bone formation in 4 and 12 weeks (Figure 3). Assessment of collagen fiber formation, an indicator of bone formation after 4 weeks, showed that the co-doped sample had the highest amount of collagen fiber formation. Most importantly, bone formation results after 12 weeks of implantation demonstrated that the Co-Sr co-doped specimens outperformed the other samples, with a bone regeneration rate of about 61%. Moreover, the strength and hardness of the new bone in contact with the co-doped specimens were higher than those of the other specimens.

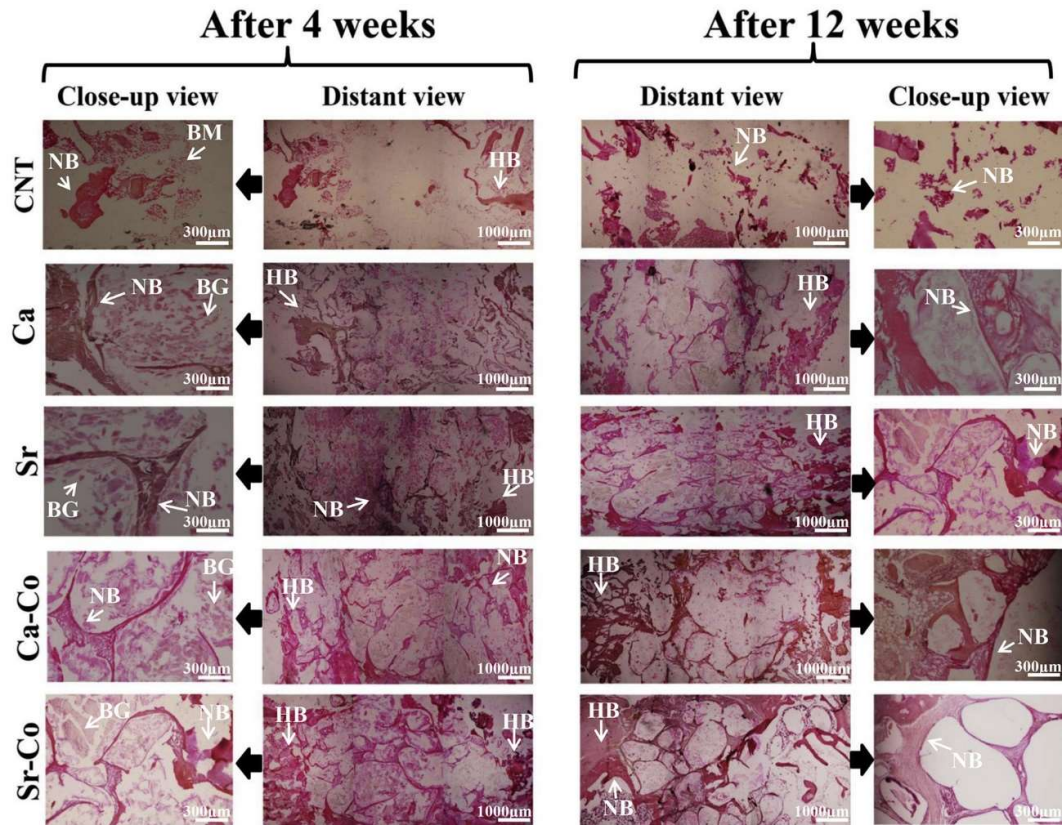

Figure 3. Micrographs of control (CNT), non-doped (Ca), Sr mono-doped (Sr), Co mono-doped (Ca-Co), and Sr-Co co-doped (Sr-Co) $SiO_2$–$P_2O_5$– CaO–CoO–SrO–$Na_2O$–$K_2O$–MgO glass samples implanted for 4 and 12 weeks in the femur of rabbits. (BM: bone marrow, BG: bioactive glass, NB: new bone, and HB: host bone), reprinted with permission from Ref. [108].

Alasvand et al. [109] studied the cell activity and apatite formation behavior of Co-Sr co-doped bioglasses with the composition of 58% $SiO_2$-(28-X)% CaO-9% $P_2O_5$-5% SrO-X% CoO (wt%) where X equals 0 or 2. The cell activity of the samples toward BMSCs indicated better cell attachment and cell proliferation on the co-doped samples, suggesting that the simultaneous addition of Co is more favorable than Sr mono-doping. Furthermore, a stronger expression of osteogenic-related genes such as the osteonectin gene and Col 1 gene was observed in the co-doped samples. Alongside their ability to form bone nodules after 21 days of immersion in SBF, the results indicated that this co-doping led to more favorable results in terms of osteogenesis compared to Sr mono-doping.

In conclusion, despite antithesis *in vitro* results on Co-Sr co-doped silicate bioceramics, likely due to differences in the composition of glasses and the content of the dopants in different researches, *in vivo* findings reported by Kargozar et al. [108] provide hope for the potential of this combination in biomedical applications, particularly in bone regeneration. However, a careful consideration and optimization of the Co concentration are crucial before implementing this combination in clinical settings.

### 3.10. Copper-Silver

Cu-Ag co-doping has been utilized in areas outside of biomedical applications, for instance, to enhance the optical and/or electrical properties of ZnO thin films [162, 163], ZnO nanostructures for the photocatalytic removal of methylene blue [164], and $TiO_2$ nanoparticles for the degradation of methyl orange under visible light [165]. In the realm of biomedical applications, this pair has been used in potassium titanate ($K_2Ti_6O_{13}$) nanowires by Lei et al. [166], improving antibacterial properties and cytocompatibility. Another work by Lysenko et al. [167] involved producing bioglass composited with co-doped BCP, indicating their ability to stimulate bone regeneration and provide antibacterial properties *in vivo*.

Munyensanga et al. [110] produced 82S5 bioglass with 0.1% mol Ag and 0, 0.1, and 0.2% mol Cu, referred to as S2P, S3P, and S4P, respectively. The results indicated that all the samples were able to form a cloud-like apatite layer after immersion in SBF. Additionally, the antifungal effect of these samples was evaluated by measuring the inhibition efficiency against Candida albicans growth. The results indicated that increasing the value of the dopants led to a higher inhibition efficiency, with the S4P samples having the best performance, followed by S3P, S2P, and then the non-doped sample.

In conclusion, the use of Ag-Cu as co-dopants can be a great choice, especially in cases where there may be bacterial attacks, as both of these dopants have antibacterial effects, as shown Munyensanga et al. [110]. In addition, it has been observed that they are able to promote apatite formation in SBF, making them a candidate for bone tissue engineering. However, the information on this pair is still not sufficient for a definitive conclusion on their abilities, and more research needs to be conducted for a better understanding of their performance. This is especially true in cases such as optimizing their content to avoid cytotoxicity and achieving the best antibacterial performance, and to obtain a thorough understanding of the interaction of Cu with the glass structure.

### 3.11. Copper-Strontium

The use of the Cu-Sr combination has been reported in non-medical applications to enhance the structural, thermoluminescence, and optical behaviors of lithium borate glass [168, 169]. In biomedical applications, the incorporation of Cu and Sr into hydroxyapatite (HA) [170], biphasic calcium phosphate (BCP) [171], HA-gelatin-Ti composite [172], and borate glass [173] has been explored. The interest in this combination stems from its potential to promote osteogenic cell activity and antibacterial activity.

Regarding biomedical silicates, Pang et al. [111] synthesized Cu-Sr co-doped diopside using robocasting. The addition of Sr and Cu-Sr co-doping gave rise to a slight decline in the grain size owing to the grain boundary pinning effect, which inhibited grain growth with improving mechanical properties. After 28 days of incubation in SBF, all the specimens displayed the precipitation of flake-like apatite. Saos-2 cell viability tests indicated higher cytocompatibility in the Sr mono-doped sample. Although the co-doped samples exhibited slightly lower cell viability compared to the Sr mono-doped sample, they still had higher viability compared to diopside (Figure 4 a-d). Cell adhesion was not negatively affected by doping with Cu and Sr, as evidenced by the presence of F-actin in all the samples. It is worth noting that cell spreading on Sr-doped diopside was slightly higher than in the co-doped sample (Figure 4 e-h). This difference in cell spreading could be attributed to the impact of doping on the lattice. The larger ionic radius of Sr than Ca leads to lattice expansion, resulting in more rapid release of Cu into the medium. This higher release of Cu can decrease cell viability due to the adverse effects of high Cu levels. The antibacterial impact of doping was assessed against *E. coli*, revealing that the Sr mono-doped sample had a slightly higher antibacterial effect compared to non-doped diopside, while the co-doped samples exhibited the highest inhibition rate of around 60% due to the additive effect of the dopants. Sr increases the release of the ions from the ceramic, while Cu produces reactive oxygen species upsetting the bacteria.

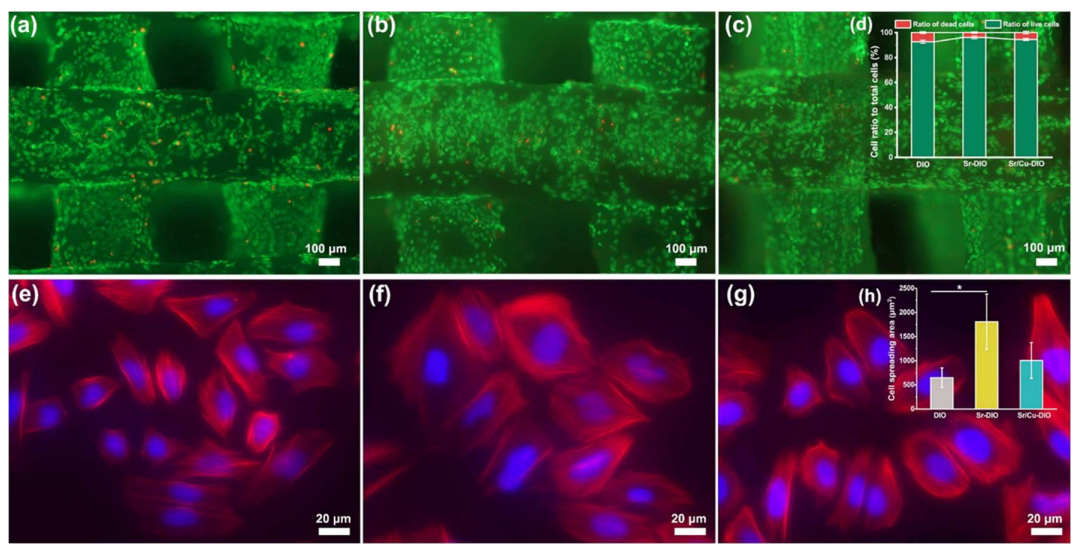

Figure 4. Fluorescence graphs of cells on non- (a), Sr- (b), and Sr-Cu co-doped (c) diopside surfaces, where green is an indicator of live cells and red is that of dead cells. Quantitative staining assay results (d). Immunofluorescence graphs of cells cultured on non- (e), Sr- (f), and Sr-Cu co-doped (g) diopside, where red

stains indicate F-actin and blue ones show nucleus. Cell spreading areas on the various samples (h), reprinted with permission from Ref. [111].

Anand et al. [112] also assessed the influence of Cu-Sr co-doping on the characteristics of mesoporous silica glasses. Cu and Sr were doped into 80% $SiO_2$-(15-2x)% CaO-5% $P_2O_5$-x% CuO-x% SrO using a sol-gel method, with x set at 0.5, 1, and 2 mol%. All the specimens indicated the precipitation of cauliflower-like apatite in SBF, but the co-dopants delayed the apatite deposition. Cell viability and antibacterial results were in agreement with Pang et al. [111] with the non-doped sample demonstrating higher cell viability (>77%) compared to the doped samples. Moreover, the specimens with the highest dopant concentration exhibited the strongest antibacterial influence against *E. coli* and *S. aureus* bacteria. Ozarslan et al. [113] also focused on the impact of Cu-Sr co-doping into 50S6P glass at 4 mol% Sr and 0.1-0.3 mol% Cu. All the specimens exhibited the deposition of cauliflower-like carbonated apatite, with the addition of Cu delaying the release of ions from the biomaterial and resulting in a delayed deposition of the apatite layer.

In another study, Yangguang et al. [114] created a bone cement from β-$Ca_{1.5}$ $Sr_{0.3}$ $Cu_{0.2}$ $SiO_4$ microsphere powder. After eight weeks of implantation in a New Zealand white rabbit root periapical bone defect model, the bone cements began to degrade. The formation of new bone tissues and blood vessels was observed, and this was superior to that of cements made from the non-doped microspheres. Additionally, no inflammatory reaction was observed. Alasvand et al. [109] also used a sol-gel method to produce Sr mono-doped (X=0) and Cu-Sr (X=2) co-doped 58% $SiO_2$-(28-X)% CaO-9% $P_2O_5$-5% SrO-X% CuO bioglasses. After 21 days of immersion in SBF, nodule bone formation was observed, indicating the bioactivity of the samples. The cytocompatibility of the samples was measured for BMSCs, showing that the co-doped samples had more cytocompatibility and cell adhesion ability compared to the Sr mono-doped samples. Moreover, it was observed that in case of the expression of osteogenic related genes, specifically Col 1 and osteonectin gene, the co-doped samples outperformed the Sr mono-doped samples. This depicts their potential application in bone regenerative applications.

Research findings on Cu-Sr co-doped silicate biomaterials, especially *in vivo* ability to promote osteogenesis and vascularization, suggest that this combination holds promise for medical applications. However, optimization is necessary to overcome the adverse effects of Cu at high doses. A trade-off exists between cell viability and antibacterial properties, which should be taken into consideration.

### 3.12. Erbium-Ytterbium

Co-doping of bioceramics like BCP with rare earth elements has been investigated for various applications, including luminescence imaging [174], drug delivery [175], and bio imaging [176]. Silicate bioceramics such as akermanite [74], CaSi [75], and Ca-Si-Ti system [115] have been explored after incorporating Er and Yb. In the context of biomedical applications, the ability of light to penetrate tissues is crucial for detection sensitivity. Red light, which has higher penetration ability than green light, is considered suitable. The incorporation of a high level of Yb has been shown to stimulate energy transfer from Yb ions to Er ions, thereby enhancing upconversion luminescence and red light intensities [74].

Li et al. [74, 75] fabricated upconversion Er- and Yb-doped akermanite glass and $CaSiO_3$ powders through a container-less processing approach. The Er-doped $CaSiO_3$ glasses exhibited low luminescence ability, but co-incorporation improved the luminescence intensity. Similar results were obtained for co-doped akermanite. The samples exhibited luminescence properties by emitting red and green up-conversion lights, with the intensity of emitted lights enhanced with the content of the dopants. However, increasing the doping level reduced the apatite formation ability of these inherently bioactive silicates, but without detrimental effects on cell cytocompatibility and proliferation, while improving angiogenic gene expression. Another research study demonstrated that Er-Yb co-doping into 70% $CaSiO_3$–30% $TiO_2$ glass systems, up to around 25%, did not influence the apatite-formation ability, whereas the apatite precipitates decreased the fluorescence intensity [115].

In conclusion, Er-Yb co-doping into silicates is an effective approach to enhance luminescence properties for bioimaging applications, with no significant impact on bioactivity and biocompatibility. Although no significant detrimental effect of this co-doping on the bone-forming ability of these silicate bioceramics has been pointed out, Er-Yb doping should only be considered in situations where the luminescence property is important and bone formation is not the primary objective.

### 3.13. Europium-Gadolinium

The investigation of materials co-doped with Eu and Gd for non-biomedical applications has been focused on improving the energy conversion of solar cells, such as $SnO_2$ [177]. Over the past decade, the utilization of rare earth elements as co-dopants for tumor imaging and therapy has been extensively studied due to their special magneto-optical properties. In this regard, Xie et al. [178] focused on the impact of co-doping on the characteristics of HA nanocrystals, demonstrating their cytocompatibility, degradability, and *in vivo* fluorescent imaging functionality. He et al. [179] also confirmed the cytocompatibility and fluorescent properties, with potential applications in cell bioimaging, resulting from Eu-Gd co-doping in calcium phosphates.

Niu et al. [116] reported a research study on the impact of Eu-Gd co-doping on the characteristics of 60% $SiO_2$-(28−x)% CaO-4% $P_2O_5$-8% Eu-x% Gd bioglass nanoparticles, where x is equal to 0, 2, 5, or 8. The co-doped nanoparticles exhibited high cytocompatibility and hemocompatibility. Moreover, the co-doped samples showcased red fluorescence emission associated with the strong fluorescence of Eu. Additionally, excellent magnetic resonance and computed tomography imaging capacities were observed.

In conclusion, Eu-Gd co-doping has the potential to yield suitable materials for bio-imaging applications, as they possess fluorescence properties without any signs of cytotoxicity. Since there is no available data on whether Eu-Gd co-doping can induce bone formation, the application of Eu-Gd co-doped silicate bioceramics should be considered solely for bio-imaging purposes. Further research is necessary to thoroughly evaluate the suitability of Eu-Gd as a doping pair for clinical applications.

### 3.14. Europium-Strontium

In non-biomedical applications, co-doping with Sr and Eu has been explored to improve the fluorescent properties of YAG phosphors [180] and the conductivity of ceria [181]. However, to our knowledge, no research has been conducted on the biomedical applications of these pairs outside the family of silicate bioceramics.

In case of silicate bioceramics, Yangguang et al. [114] patented a type of co-doped silicate bioceramics with investigating their biological properties. They produced β-$Ca_{1.7}$ $Eu_{0.1}$ $Sr_{0.2}$ $SiO_4$ granular powder and used it as raw material to produce bone cements. The results of compression tests indicated that the addition of Eu and Sr as dopants led to a 20% increase in compression strength, up to 55 MPa, compared to the non-doped bone cements. The *in vivo* tests of the bone cements in an adult male New Zealand white rabbit femur defect model indicated that the cement started to degrade after four weeks, in addition to the adhesion of osteoblast cells and new bone formation. After 12 weeks, the new bone became compact, and the formation of small blood vessels was observed, indicating the great potential of these doped silicate bioceramics as a bone regenerative material. Using the same raw powder, they also created another cement to explore their drug release ability for dental defects. The gentamicin drug release in SBF indicated a rapid release was maintained for the first 10 days by keeping the concentration high. Then, after 21 days, the release became uniform. An *in vivo* test on a chronic periapical bone defect indicated degradation after eight weeks, along with new bone tissue formation and micro-angiogenesis.

The results from the patent indicate that bone cements obtained from CaSi powders doped with Eu and Sr can exhibit both angiogenesis and osteogenesis, which is crucial for an effective bone regenerative device. Also, a desired release behavior in SBF shows a promising application of this combination as an integrated drug delivery and tissue engineering device.

### 3.15. Fluorine-Strontium

The co-incorporation of F and Sr has been explored by Jumpatam et al. [182] to enhance the dielectric behavior of $CaCu_3Ti_4O_{12}$. In the field of medical applications, Huang et al. [183] demonstrated the superior corrosion protection and cytocompatibility of F-Sr co-doped HA coatings on titanium surfaces compared to non-doped HA coatings. Additionally, Pourreza et al. [184] indicated the higher cytocompatibility of F-Sr co-doped BCPs compared to the non-doped samples.

Regarding silicate bioceramics, Shahrouzifar et al. [117] evaluated the impact of F and Sr co-doping on the bio-performance of diopside scaffolds. While F and Sr mono-doping had a positive influence, co-doping negatively affected the apatite-formation ability due to the high affinity of these two ions, limiting bioresorption. The dopants also influenced the morphology of the deposited apatites. Based on Figure 5, the non-doped scaffolds exhibited plate-like apatite, while the mono-doped samples showed spherical apatite deposition. The co-doped samples displayed a combination of spherical and plate-like apatites. The introduction of F reduced the cytocompatibility of the bioceramic, whereas Sr-doped samples exhibited the most improved bio-performance.

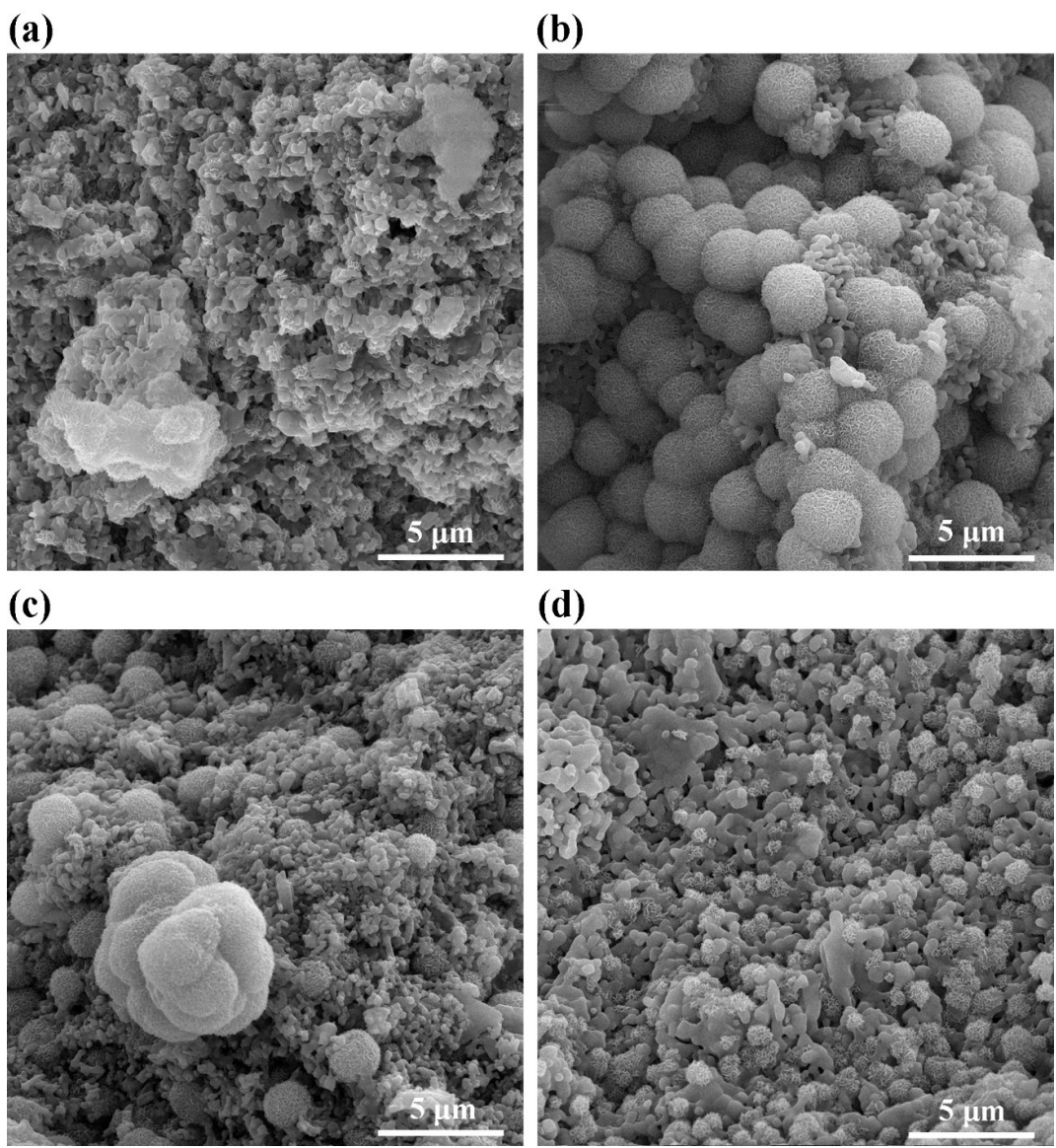

Figure 5. Microscopic images of non- (a), F- (b), Sr- (c), and F-Sr co-doped (d) diopside samples after 7 days of incubation in the physiological medium, reprinted with permission from Ref. [117].

Although F and Sr individually exhibit positive effects on bone formation, they are not suitable candidates for co-doping of biomedical silicates as they reduce biodegradability, bioactivity, and biocompatibility due to a destructive interaction.

### 3.16. Lithium-Silver

Co-doping of Li and Ag into NiO has been investigated to improve solar cell performances [185]. In medical applications, both Li and Ag are known for their antibacterial properties; hence, their simultaneous use is expected to impart favorable antibacterial properties to bioceramics.

Regarding silicates, Rahmani et al. [118] synthesized Li-Ag-doped 58S bioglasses with a constant Li concentration of 5% and the varying amount of Ag (0, 1, 5, and 10%). The glasses containing 1% Ag exhibited optimized behaviors in terms of degradability, apatite-formation ability, antibacterial efficiency, MC3T3-E1 cell response, and ALP activity. In another work, the influence of Li-Ag co-doping into 68S glass was assessed at a concentration of 5 mol% for both [119]. The doped samples demonstrated the formation of sphere-shaped HA, no cytotoxicity to MC3T3-E1 cells, induced ALP activity, and a significant bactericidal impact against *E. coli* bacteria. These properties were superior to those of the non-doped sample, due to the combined impact of Ag and Li. Also, Li-Ag co-doped 58S and 68S glasses, with the constant amount of Li at 5% and the varying amount of Ag at 0, 5, and 10% was analyzed [120]. Th apatite-formation ability was not adversely affected in both composition by Li-Ag doping. Cell activity tests indicated that the samples doped with 5% Ag, both in case of 58S and 68S glasses, had the best performance compared to Li mono-doped and 10% Ag-doped samples. This is because the addition of Ag to the structure leads to a more porous structure, providing more surface area and stimulated cell activity. On the other hand, the large value of Ag can be cytotoxic, affecting the performance of 10% Ag-doped samples. Between the two cases of 5% Ag-doped samples, however, 58S glass performed better compared to 68S. Also, the 5% Ag-doped samples had the optimal antibacterial property against MRSA bacteria.

In conclusion, the properties of Li-Ag co-doped silicates depend on the dopant levels, emphasizing the need for an optimization approach to achieve improved bio-performance. Considering the remarkable bactericidal ability of Ag and the capability of Li to stimulate cell proliferation and osteogenesis-related cell expression, the combination of Li and Ag holds promise for bone regenerative applications. However, since both elements can potentially induce cytotoxicity, further research is necessary to determine the optimal composition.

### 3.17. Magnesium-Silver

Co-incorporation of Mg and Ag has been investigated in bismuth strontium calcium copper oxide for electrical applications [186]. Additionally, this combination has been studied in medical-grade HA [187] and Ti-6Al-4V [188]. In case of this co-doping pair, Ag contributes to antibacterial properties, while Mg enhances mechanical properties and counteracts the negative effects of Ag on cell viability.

Kaur et al. [121] pointed out the co-application of Mg and Ag in 35.5% $SiO_2$–4% $P_2O_5$–40% CaO–20.5% $Na_2O$ nanoparticles. HA was observed on the nanoparticles after immersion in a physiological environment. However, the amount of Mg and Ag incorporated in the nanoparticles limited the precipitation of HA owing to the impact of Mg on the stability of the glasses. Mg ions enter the silicate network, increasing interconnectivity by introducing stable

Mg-O bonds with a relatively high covalent character. The optimal amount of Ag, leading to the complete inhibition of bacterial growth, was measured at 0.75 mg/ml (Figure 6). The nanoparticles exhibited more than 80% cell viability, but the proliferation of MG63 cells was declined by enhancing the Ag level owing to the cytotoxic nature of Ag ions. However, even co-doping of Mg and Ag showed potential for antibacterial applications, further optimization is required to balance the antibacterial effect and cytocompatibility.

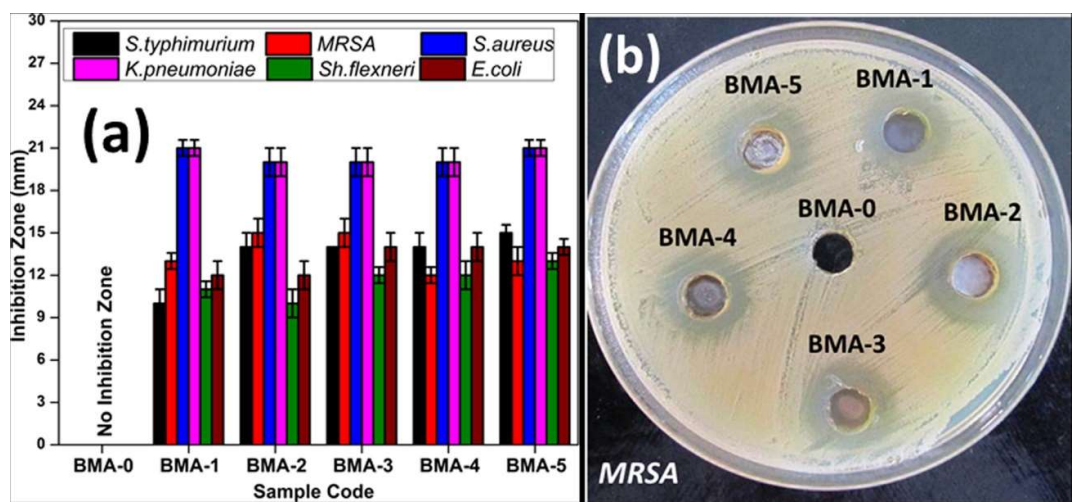

Figure 6. Amounts of the inhibition zone of different types of bacteria exposed to Mg-Ag doped CaO–Na$_2$O–SiO$_2$–P$_2$O$_5$ samples (a) and macrograph of bacterial activity (b) (the samples with the BMA-x code contain 2x mol% of each dopant), reprinted with permission from Ref. [121].

Ebrahimi et al. [122] conducted a study on Ag-Mg co-doped 58S glasses with a composition of 60% SiO$_2$–26% CaO–5% SrO-4% P$_2$O$_5$-x% MgO-(5-x)% Ag$_2$O (with x ranging from 0 to 5). The samples exhibited apatite formation ability; however, the addition of the dopants resulted in delayed apatite formation and morphological changes. Viability tests on G292 osteoblast cells showed that samples with a low amount of Ag did not exhibit cytotoxicity. In contrast, the Ag mono-doped sample and the sample with 4 mol% Ag and 1 mol% Mg showed lower cell viability compared to the non-doped sample. However, the specimens with a higher amount of Ag depicted outstanding bactericidal influences against *E. coli* and *S. aureus* bacteria. Ultimately, the sample containing 2 mol% Ag and 3 mol% Mg was identified as the optimized sample.

In conclusion, co-doping of silicate bioceramics with Ag and Mg involves a trade-off between properties. Ag provides antibacterial ability but reduces cell viability and proliferation. The addition of Mg can help balance these side effects, although it may decrease apatite formation. Ag-Mg co-doped ceramics show promise for bone regeneration applications as they exhibit osteogenesis ability and antibacterial behavior. Further research and optimization are required to determine the ideal composition for specific applications.

### 3.18. Magnesium-Strontium

Co-doping of Mg and Sr in various materials has shown promise in different fields, including electrical and biomedical applications. In electrical applications, co-doping of Mg and Sr has been explored in lanthanum gallate [189]. This combination has also been used for structure manipulation purposes [190]. In biomedical applications, the co-incorporation of Mg and Sr into materials like HA [191], fluorapatite [192], and tricalcium phosphate (TCP) [193] has been investigated. This co-doping pair has been proven to enhance the stability of amorphous calcium phosphate (ACP) and inhibit the transition of this phase to HA. It is achieved through the synergistic influence of Mg and Sr, which alters the lattice parameter of apatite. Mg and Sr are both known for their ability to promote bone formation, making them desirable dopants in bioceramics.

Wang et al. [123] focused on Mg-Sr co-doping into wollastonite. Although it has been proven that the incorporation of Mg into wollastonite can considerably enhance compressive strength [26], Sr-Mg co-doped wollastonite exhibited lower strength than Mg-doped wollastonite, typically 56 MPa vs. 92 MPa. It is also noticeable that Mg incorporation delays the degradability and apatite-formation ability of wollastonite, but the co-doped wollastonite exhibited enhanced apatite formation ability and accelerated degradation compared to Mg mono-doped wollastonite. Additionally, the co-doped samples indicated comparatively higher cell activities. In another report [124], the high cell viability and ability to proliferate human fibroblasts by Mg-Sr-doped hardystonite have been confirmed. Furthermore, Rezaei et al. [125] confirmed the apatite formation ability, cell biocompatibility, proliferation, and ALP activity of Mg-Sr co-doped 45S5 bioglass. In another study, Li et al. [126] demonstrated that Mg-Sr co-doped wollastonite was able to promote both angiogenesis and osteogenesis. This was observed *in vitro* and *in vivo*, the latterin femoral-defect model rabbits. However, their primary focus was on the impact of pore geometry on these phenomena, so they did not provide further information about the effects of co-doping.

It is accordingly concluded that co-doping of Mg and Sr into silicates appears to be more promising than their mono-doped counterparts, as each element compensates for the weaknesses of the other. Additionally, since both Mg and Sr have well-established benefits in promoting bone formation both *in vitro* and *in vivo*, this duo has great potential for bone regenerative applications. However, the interactions of Mg with Sr and the ceramic matrix are complex, and further experiments and research are needed to gain a comprehensive understanding of how this co-doping pair affects bioceramics from various aspects. Such knowledge is crucial before advancing toward clinical applications in the field of bone regeneration.

### 3.19. Magnesium-Zinc

The utilization of Mg-Zn co-doping, as well as the biomedical field, has been investigated in the optoelectronic field for NiO [194]. In biomedical applications, Mg is recognized to be essential for promoting bone formation, whereas Zn not only increases bone-related cell activity but also has an inhibitory effect on bone bioresorption by constraining the activity of osteoclasts. Additionally, Zn has an antibacterial effect against certain bacteria. Zhao et al. [195] fabricated Zn-Mg co-doped $TiO_2$ films on implants, leading to improved cell activity and osteogenesis ability.

Abdalla et al. [127] fabricated CaSi cements doped with Mg and Zn, with varying Ca/Zn ratios. The increase in the Zn content reduced the apatite-formation ability of the cement owing to the retarding effect of Zn. Zn stabilizes the network, controls the release of Ca, and is adsorbed on active sites of HA, constraining the nucleation of HA [196]. However, a high level of Zn incorporation adversely affects fibroblast cytocompatibility due to the cytotoxic impact of Zn at higher concentrations. The decrease in apatite-formation ability as a result of Zn incorporation in both mono- and Mg-Zn co-doping cases was also confirmed by Erol et al. [128].

Vikas Anand et al. [129] investigated the characteristics of 46.1% $SiO_2$-26.9% CaO-2% MgO-2.6% $P_2O_5$-x% ZnO-(22.4-x)% $Na_2O$ glass with x = 6, 8, and 10. The results of immersing the specimens in SBF displayed that all of the specimens were able capable of growing apatite on the surface. Nevertheless, the growth rate of apatite decreases by increasing the Zn content due to two reasons. First, both Mg and Zn can play the role of either network former or modifier. Based on this dual behavior, they can affect the structure of the silicate bioceramics differently. When acting as network formers, the structure weakens as the newly formed bonds are weaker than the existing bonds. However, when functioning as network modifiers, they strengthen the structure, leading to lower degradation and consequently reduced ion release. Second, the decrease in the apatite growth rate is attributed to the reduced surface area available for apatite growth as the Zn content increases. Based on BET results, the pore size decreases with increasing the Zn content, which is again another result of Mg and Zn acting as network modifiers. Furthermore, the gentamicin drug release behavior of the specimens was investigated, revealing similar release profiles with the other specimens.

It can be concluded that the co-incorporation of bioceramics with Mg and Zn may not be beneficial for bone regenerative applications. On the one hand, degradation is strongly retarded, leading to reduced apatite-formation ability. On the other hand, exceeding the safe threshold for the Zn content results in noticeable cytotoxicity. Therefore, Mg mono-doping appears to be more advantageous. However, considering that the performance of these dopants depends heavily on whether they act as network modifiers or network formers, it is possible that this combination can have more favorable effects on other types of silicate bioceramics, which warrants further research.

### 3.20. Manganese-Zinc

Mn-Zn co-doping of $BiFeO_3$ and $Fe_3O_4$ with the aim of altering their magnetic properties for various applications, such as hyperthermia, has been reported [197, 198]. Mn plays an essential role in the creation of the extracellular matrix and proteins like collagen, while Zn promotes the proliferation of osteoblasts and provides antibacterial effects.

Sarin et al. [130] assessed the impact of Mn-Zn co-doping on the bio-performance of 43% CaO–42% $SiO_2$–9% $P_2O_5$–6% ZnO bioceramics with a constant content of 6% Zn and varying levels of Mn. High levels of Mn (beyond 6%) reduced the apatite-formation ability caused by Zn doping. This effect may be due to the disruption of stable bridging oxygen units by Mn, leading to increased degradation and subsequent release of ions. Furthermore, Mn has been shown to enhance the cell viability of the specimens, with the highest level of Mn exhibiting a cell viability of approximately 92%.

In conclusion, Mn-Zn co-doping shows promise for bone regenerative applications as it improves biomineralization and cytocompatibility while retaining the beneficial features of Zn such as antibacterial and osteoblast activities. However, further research is needed to evaluate its impact on gene expression, antibacterial effects, *in vivo* bone formation ability, and other factors.

### 3.21. Selenium-Strontium

The co-doping of Se and Sr into bioceramics has been investigated for HA [199] due to the antibacterial effect of Se and the bone formation-inducing ability of Sr. In case of silicate bioceramics, Sarin et al. [131, 132] investigated the simultaneous incorporation of Sr and Se into 43% CaO–40% $SiO_2$–12% $P_2O_5$–5% MgO bioceramics. The co-doped samples exhibited higher degradability and HA formation due to the presence of Sr, as well as enhanced antibacterial performance attributed to both dopants. Regarding cell viability, Sr had a helpful contribution, while Se inhibited the proliferation of cancerous MG63 cells but enhanced MC3T3-E1 adhesion, proliferation, and differentiation.

In conclusion, Se-Sr co-doping in silicates offers improved bioresorbability, bioactivity, and antibacterial properties. However, Se exhibits cytotoxicity to certain cell types like MC3T3-E1. Therefore, due to insufficient data, it is not possible to determine whether this combination holds promise for future clinical applications. Further research and data are necessary to better understand the potential of this co-doping pair.

### *3.22. Silver-Strontium*

To our knowledge, the use of Ag-Sr co-doping in non-medical applications has not been reported. However, in medical applications, Ag-Sr co-doping of HA [200] and titanium phosphate [201] has been explored. Sr is commonly used as a dopant in bioceramics to promote bone generation, while Ag is known for its antibacterial activity. It is important to note that excessive Ag dosages cause highly cytotoxic effects. The intriguing aspect of this co-incorporation is the ability of Sr to delay the release kinetics of Ag ions and regulate their cytotoxic effects. Studies have also shown that released Ag ions preferentially bind to ALP sites, leading to functional destabilization. However, these sites have demonstrated a greater affinity for Sr or other species like Mg or Zn, rather than Ag [200].

Swe et al. [133] doped S53P4 bioglass containing 3 mol% Sr and 1 mol% Ag. Analysis of HA formation on the samples soaked in the Hank's solution showed that the co-doped sample had superior HA formation ability compared to the non-doped or mono-doped ones. Incorporating Ag into the Sr-doped glasses transformed the semi-crystalline structure to fully crystalline, mainly due to the substitution of AgO with $Na_2O$, which facilitated faster HA formation. Additionally, the co-doped specimen depicted a higher release rate of Sr and a lower release rate of Ag compared to the mono-doped specimen, confirming the inhibitory role of Sr incorporation on Ag release. In terms of cell activity, the co-doped samples showed superior cell viability and proliferation compared to the non-doped and mono-doped specimens. The formation of bone-like nodules after MC3T3-E1 cells culture for 28 days and staining with Alizarin Red was also observed (Figure 7). The results showed that stain distribution on the non-doped samples was non-uniform, whereas the addition of Sr and/or Ag improved the uniformity of stain distribution. When considering the bactericidal properties, the Ag mono-doped sample exhibited the highest antibacterial ability, with a negligible difference compared to the co-doped sample. Another study conducted on mesoporous bioglass nanoparticles with a composition of 70% $SiO_2$–24% CaO–5% SrO–1% AgO confirmed a decrease in Ag release when combined with Sr incorporation. This effect was limited by the high specific surface areas of the nanoparticles [83]. In a patent, Yangguang et al. [114] introduced a type of bone cement composed of β-$Ca_{1.85}$ $Sr_{0.1}$ $Ag_{0.05}$ $SiO_4$ granular powder. The bone cement derived from the co-doped powders demonstrated a compression strength of 75±15 MPa. This represents a significant 63% increase in strength compared to cement fabricated from non-doped powders. Furthermore, *in vivo* testing conducted on a skull defect model of a healthy adult male New Zealand white rabbit revealed promising results. After a period of 12 months post-implantation, 65% of the cement was replaced by new bone growth with small blood vessels formed in the implantation site.

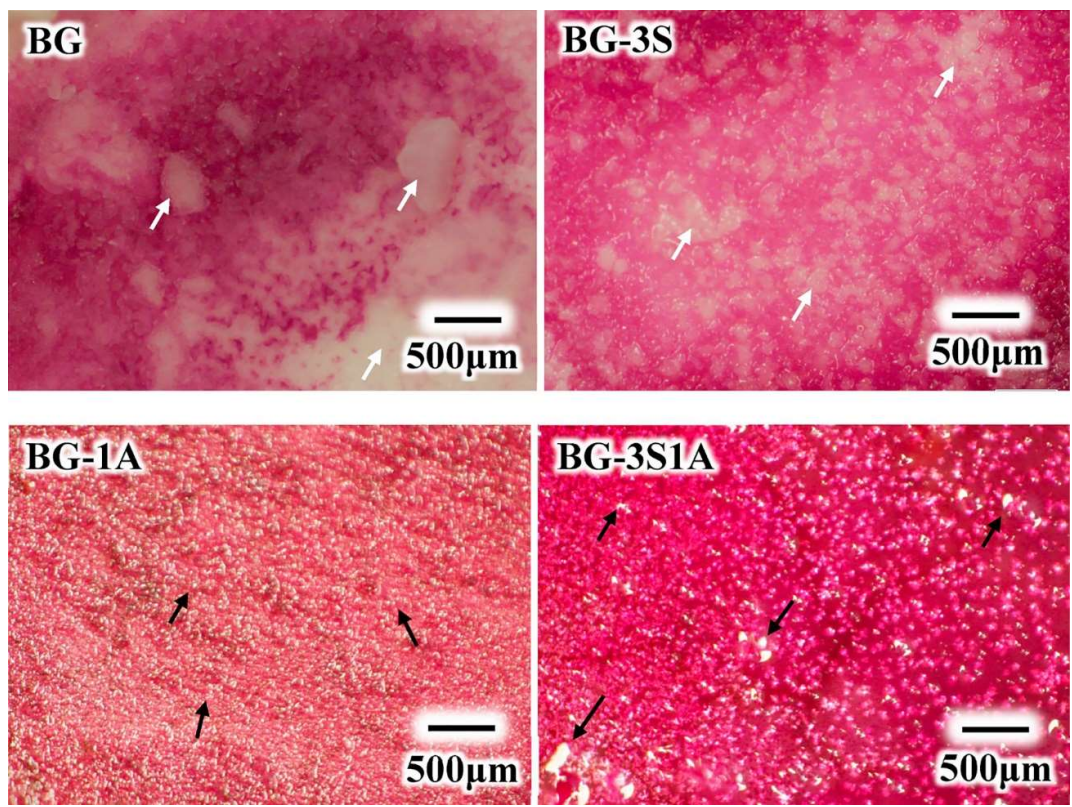

Figure 7. Formation of bone-like nodules on S53P4 bioglass co-doped with Sr and Ag after 28 days of MC3T3-E1 cell culture (black and white arrows are silver particles and unstained area, respectively), reprinted with permission from Ref. [133].

In conclusion, co-doping of Ag and Sr is an effective approach for enhancing apatite formation ability and cell activity in silicates, although it results in a slight decrease in antibacterial performance compared to Ag mono-doped states. The remarkable bactericidal properties of Ag, combined with the bone regeneration promotion ability of Sr, indicate that these two dopants complement each other and offer a promising composition for bone treatment. Further research is needed to optimize the dosage of Ag, through *in vivo* and clinical tests, to fully assess the effectiveness of this combination for bone regeneration.

### 3.23. Silver-Zinc

Co-doping of Ag and Zn in non-biomedical applications has been reported to modify the optoelectronic properties of CdO thin films [202] and photocatalytic behavior of TiO$_2$ nanoparticles [203]. In biomedical applications, the co-incorporation of Ag and Zn is anticipated to provide bone regenerative and antibacterial effects. The possibility of co-doping of biomaterials with Ag and Zn has been investigated for titanium [204], β-TCP [205], and HA [206]. Regarding silicate bioceramics, Ningsih et al. [134] fabricated 76S bioglass with 2.5 mol% of Ag and Zn. For comparison, 2.5 and 5 mol% Ag mono-doped samples were also investigated. The antibacterial effect of bioglasses against *E. coli* bacteria was assessed where the antibacterial rate for the non-doped, 2.5% Ag mono-doped, 5% Ag mono-doped, and co-doped samples were 7.8%, 55.1%, 99.3%, and 100%, respectively. These results indicate the proper antibacterial effect of co-doping with Ag and Zn. However, in case of apatite formation ability in SBF, the order was reversed, with the co-doped sample being the worst and the non-doped sample being the best. This can be mostly due to the effect of Zn on delaying the apatite formation. However, to some extent, all the samples were able to support the *in vitro* apatite formation. Cell viability testing toward MC3T3-E1 indicated that all the samples except for the 5% Ag mono-doped bioglass were cytocompatible, where the co-doped sample was slightly better than the others, especially after 72 h. Chetan and Vijayalakshmi [135] investigated the co-doping effect of Ag and Zn into strontium phosphrosilicate (Sr$_5$(PO$_4$)$_2$SiO$_4$). Sr in the composition was substituted by the dopants, where four co-doped samples were produced in a way that the molar composition of doped glasses was as (1-2X)% Sr-0.4% PO$_4$-0.2% SiO$_4$-X% ZnO-X% AgO (X=0-0.04), called 1x, 2x, 3x, and 4x, respectively. Hemolysis assessments indicated that the addition of the co-dopants up to 3x did not damage the structural integrity of the red blood cells, whereas in case of 4x substitution, red blood cells were clearly damaged. The cytotoxicity tests of the samples in an acidic medium toward MG-63 cells indicated that the non-doped samples were the most cytocompatible. However, the 2x samples, only at very low concentrations (25 μg/ml), showcased slightly better results. But at higher concentrations, even 2x samples had an inferior performance. It is notable that the 4x samples were cytotoxic as a drastic decrease in cell density alongside ruptured cells were observable. Considering the antibacterial effect of the co-doped samples, they were effectively able to act against *E. coli* bacteria. However, *S. aureus* bacteria were able to resist the antibacterial medium due to the presence of a thick peptidoglycan layer. SBF studies results also indicated the ability of particles to form ACP flakes. At the end, based on the results, they suggested that the samples with 6% co-doping, being 3% of each dopant, were most effective in the promotion of osteogenesis and antibacterial effects.

Naruphontjirakul et al. [136] fabricated bioactive glass with a composition of 46.48% $SiO_2$-17.52% CaO-35.97% SrO with 0, 0.5, 1, and 1.5 mol% Ag and 0, 1, and 1.5 mol% Zn. After 21 days of immersion in SBF, cauliflower-like apatite formation on the samples was observed, indicating their probable ability to bond to the host bone. The cell viability toward MC3T3-E1 cells indicated that the Zn-doped samples had no cytotoxicity effect until a concentration of 250 μg/mL. However, the increase in the concentration of the samples led to a decrease in the cell viability. To investigate the ability of this doping pair against cancer cells, MG-63 and hFOB 1.19 were used, where MG-63 are bone cancer cells and hFOB 1.19 are non-tumor bone cells. Regarding hFOB 1.19 cells, the particles doped with 1.5% of each Ag and Zn were non-cytotoxic up to a concentration of 125 μg/mL. At higher concentrations, the cytocompatibility of the co-doped samples significantly decreased compared to the non-doped samples, indicating the doped samples being more cytotoxic toward hFOB compared to MC3T3-E1 cells. However, the Zn mono-doped and all co-doped samples were able to effectively kill MG-63 cells at concentrations higher than 125 μg/mL, while maintaining the cell viability toward the hFOB 1.19 cells. This is because Zn is released more when in contact with an acidic environment, which is usually the case of surrounding cancer cells. It is notable that due to a more compact structure, the sample doped with 1.5% of each dopant was the least effective one. All the samples containing Zn up to 1 mol% were also able to effectively promote the differentiation of osteogenic cells due to the release of Ca, Sr, and Zn. The addition of Ag and the increase in the content of Ag caused a decrease in mineralization and cell differentiation ability of the samples. The antibacterial effect of the samples against *B. subtilis*, *S. aureus*, and *E. coli* indicated that the non-doped sample had no inhibitory effect against the bacteria. The Zn mono-doped samples only had an inhibitory effect against *B. subtilis*. However, the addition of Ag led to an antibacterial effect against all the three types of bacteria.

In conclusion, the combination of these dopants presents several challenges as a tissue engineering device. Both dopants need to meet a certain threshold to be cytocompatible, which limits the research for optimizing biological effects. Additionally, the reduction of mineralization due to the decreased degradability by Zn is another hurdle to overcome, as Ag is not able to compensate for this reduction. Therefore, the application of this pair for tissue engineering applications does not seem to hold much promise. However, their ability to kill bone cancer cells while being cytocompatible to other cells can draw attention to this pair for the treatment of bone cancer. Nevertheless, more research needs to be conducted to fully uncover their potential for the treatment of bone cancer.

### 3.24. Silver-Zirconium

The co-incorporation of Ag and Zr in ZnO [207] and $TiO_2$ [208] has been investigated in photocatalyst applications. This co-incorporation into BCP bioceramics has been investigated by Mabrouk et al. [209], resulting in a suitable apatite formation ability, cell viability, and 100% reduction of bacterial strain. However, the deterioration of mechanical properties was observed. In this pair, Ag possesses unique antibacterial characteristics, while Zr can contribute to antibacterial ability and promotion of bone regeneration by stimulating bone-related cell activities.

Zohourfazeli et al. [137] assessed the influence of Ag-Zr co-doping into 58S glass fabricated by a sol-gel method, with a fixed Zr content of 5% and varying Ag contents of 0, 2, 4, 6, and 8%. After 7 days, the deposition of HCA was observed on all of the samples, with the sample containing 4% Ag having the highest quantity of HCA precipitation. It is owing to the decrease of non-bridging oxygen bonds due to Ag incorporation and the strengthening effect of Zr. More typically, the co-incorporation of Ag and Zr led to a stronger antibacterial effect and improved MC3T3-E1 cell activity. Again, the sample with 4% Ag demonstrated the optimum biological attributes in medical applications since the higher amounts of Ag have adverse impacts upon cell viability, proliferation, and ALP activity because of the development of Ag layers. That is, the incorporation of Zr can improve the biocompatibility of Ag-doped silicates, mitigating the cytotoxicity of Ag while maintaining apatite-formation and antibacterial abilities.

However, it should be noted that neither Zr nor Ag significantly enhances osteogenesis properties. Therefore, it can be concluded that there are more favorable combinations of dopants that can provide higher osteogenesis with antibacterial effects. As a result, the Zr-Ag pair may not be a priority for future clinical applications, and further exploration of alternative dopant combinations is warranted.

### 3.25. Strontium-Zinc

The simultaneous incorporation of Sr and Zn has been assessed in materials employed in electrical and solar cell applications, such as $LaGaO_3$ [210], calcium copper titanate ($CaCu_3Ti_4O_{12}$) [211], and $TiO_2$ [212]. The co-addition of Sr and Zn into biomaterials has been investigated for various materials, including HA [213], Mg alloys [214], TCP [215], and calcium polyphosphate/graphene carbon nanosheets/ultra-high molecular weight polyethylene composites [216]. Sr is incorporated to promote bone regeneration, whereas Zn is used to further improve bone formation by inducing bone-related cell proliferation and differentiation, along with exhibiting an antibacterial effect against certain bacteria.

Zhang et al. [138] analyzed the properties of Sr-Zn co-doped akermanite. They found that a wide concentration range of the co-doped sample's extractions showed no sign of cytotoxicity and improved the ALP activity of BMSCs than the non-doped specimen. Echezarreta-Lopez et al. [139] fabricated Sr-Zn co-doped 49.46% $SiO_2$–27.27% CaO–6.60% $Na_2O$–6.60% $K_2O$–1.07% $P_2O_5$–3% MgO–3% SrO–3% ZnO nanofibers to study cytocompatibility and antibacterial properties. The co-doped samples exhibited over 90% cytocompatibility with fibroblast (BALB/3T3) cells. Moreover, high concentrations of the co-doped nanofibers in Dulbecco's Modified Eagle Medium (DMEM) (75 mg/ml) displayed antibacterial effects after the second day of soaking, whereas no such effect was observed at lower concentrations. This two-day delay can be attributed to the establishment of bacteria resistance against antibacterial agents at non-sufficient concentrations, particularly for Zn. Kapoor et al. [140] focused on the bio-performance of Sr-Zn co-doped glassy phosphosilicate. They found that by enhancing the content of the dopants, the apatite formation ability noticeably decreased, likely due to reduced degradability. Also, MG63 cell viability and proliferation increased by incorporating 2% of each dopant, while these attributes decreased with further increases in the dopant concentration due to oxidative stress damage to osteoblasts. It was eventually concluded that optimized levels of Sr-Zn co-doping into silicates improve cell and ALP activities with a moderate antibacterial property, whereas higher addition levels deteriorate both cell cytocompatibility and proliferation due to the reduction of Mg levels.

Yu et al. [217] showed that BMSCs in contact with low concentrations of Zn exhibited enhanced cell activity. Because Zn at low concentrations regulates the expression of integrins α1 and β1 and increases the intracellular content of $Zn^{2+}$ by the alteration of ZIP1 and ZnT1 release. It gives rise to the activation of signaling pathways and subsequent improvement in the release of osteogenesis-related genes. However, excessive concentrations of Zn lead to down-regulating integrins. On top of that, over-upregulating ZIP1 expression leads to excessive intracellular Zn, which inhibits BMSC differentiation and increases reactive oxygen species production accompanied by BMSC apoptosis. In another work, Yangguang et al. [114] fabricated bone cements made of $\beta$-$Ca_{1.6}$ $Sr_{0.3}$ $Zn_{0.1}$ $SiO_4$ granular powder. The implantation of the bone cement in the femur defect of healthy adult male beagle dogs and the chronic periapical bone defect model of New Zealand white rabbits indicated the formation of new bone and blood vessels without any inflammatory reaction in 12 months and 12 weeks, respectively.

Lu et al. [141] fabricated Sr-Zn co-doped wollastonite, comparing their properties with non-doped and mono-doped samples. Biomineralization tests in SBF indicated that despite the fact that both non-doped and mono-doped samples were able to produce an apatite layer after 14 days of immersion in SBF, the co-doped sample showed no mineralized products. This can be due to the synergetic effect of Zn and Sr on prohibiting the release of Ca and Si from the bioceramic, being in accordance with the results obtained from work of Kapoor et al. [140].

Cell activity results for mBMSCs indicated that the proliferation of these cells increased significantly in the Zn mono-doped and Sr-Zn co-doped samples after 3 days. However, after 5 days, despite them still promoting cell proliferation, there was no significant difference with the non-doped sample. Nevertheless, the co-doped sample exhibited the highest cytocompatibility among all the samples. Considering ALP activity and up-regulation of osteogenesis and angiogenesis related genes, the results indicated that Sr mono-doped and Sr-Zn co-doped samples had better performance compared to the other samples, especially after 7 days of immersion in SBF. This in total leads to the conclusion that in an additive way, Zn can increase the cell activity, whereas Sr improves osteogenesis ability. In addition, it has been shown that the co-doped samples were effectively able to inhibit differentiation of osteoclasts.

In conclusion, considering the ability of Sr to stimulate bone formation and the promotion of osteogenesis-related cell activity by Zn, the combination of Sr and Zn is a potential candidate for bone reconstruction for clinical applications as they have been proven to be effective in promoting bone formation also *in vivo*. Considering the already existing research and *in vivo* results on bone formation ability, with more research and attention, this pair can get closer and closer to commercialization.

### 3.26. Zinc-Zirconium

The simultaneous employment of Zn and Zr in non-biomedical applications has shown to improve the magnetic and dielectric properties of certain materials, such as yttrium iron garnet [218], barium hexaferrite [219], and $BaZnxZrxFe_{12}$-$2xO_{19}$ [220]. In the biomedical field, the addition of Zn-Zr has been investigated in hyperthermia biomaterials, such as $CoFe_2O_4$ [221] and strontium hexaferrite [222]. Of considerable properties of Zn is its potential to enhance bone formation and induce the antibacterial effect, while Zr can result is enhancements in mechanical and antibacterial properties. Therefore, co-doping with Zn and Zr is expected to offer the bactericidal effect alongside with enhanced bone regeneration.

Regarding silicates, Moghanian et al. [142] assessed the impact of Zr-Zn co-doping into 60% $SiO_2$-(31-x)% CaO-4% $P_2O_5$-5% $ZrO_2$-xZnO bioglass, where x represents varying Zn contents (0, 2, 4, and 6 mol%). Formation of globular HA on all the specimen surfaces was verified through various experiments. The addition of Zn, compared to the Zn-free sample, gave rise to enhanced cell viability and ALP activity in all of the specimens. However, the samples with 2 and 4 mol% Zn showed the best results, whereas the specimen with 6 mol% Zn exhibited the lowest cellular activity among the Zn-doped specimens. These results indicate a dose-dependent positive effect of Zn on cellular activity. As previously stated, Zn additions exceeding 10 mol% show adverse effects on biological attributes; however, in this case, the adverse effect was observed at 6% Zn doping. It suggests that the presence of Zr reduces the safe amount of Zn. To better understand the reason behind these phenomena, further studies on different factors are required. In case of antibacterial effects against MRSA bacteria, all of the samples showcased better properties compared to the Zn-free sample. However, in this case, the sample with 4 mol% Zn had the superior performance and the sample with 6 mol% still had the lowest performance.

Results obtained from Zr-Zn co-doped glasses indicated that the Zr-Zn doping duo is promising for medical applications. However, the addition of Zr reduces the safe amount of Zn, which causes adverse effects on biological properties. Therefore, further research is necessary to understand the interaction between Zr and Zn before drawing confident conclusions. Additional investigations are needed to establish the clinical viability of this co-doping duo with certainty.

## 4. Multiple co-dopants used in biomedical silicates

In the previous section, the effects of dual co-doping into silicate bioceramics were reviewed, demonstrating wondrous effects on biological and mechanical properties. However, there is still ample opportunity for further enhancing the performance of silicates through the incorporation of multiple ions. Despite this potential, there are relatively few reports on multiple co-doping into silicate bioceramics, while multiple co-doping of other bioceramics such as HA, with elements like Mg, Si, Zn, and Cu [223], Ag, Sr, Mn, and Zn [224], and Zn, Mg, and Sr [225], is further well-established.

Regarding silicates, Lázaro et al. [226] evaluated the impact of Zn-Mg-Sr doping into 64.42% $SiO_2$-24.43% CaO-5.20% $Na_2O$-5.95% $P_2O_5$ (wt%) bioglass. The incorporation of Sr, Mg, and Zn gave rise to a decline in the crystallization temperature ($T_c$), especially with Mg. The lower dissociation energy of Mg-O bonds compared with Ca-O bonds significantly affected $T_c$ when 1.95 mol% of Mg was added. Also, even a small addition of 0.01 mol% Zn played a significant role in the reduction of $T_c$, whereas Sr had no considerable effect. It was also explored that impregnation with these dopants had no extensive impact on pH changes in McCoy's 5A medium since fast ionic exchanges between the ceramic and the physiological fluid are the main reason for changes in pH and the concentration of Na is the same in all of the specimens. Also, the deposition of carbonated apatite on all of the specimens was detected after 7 days of incubation in the medium.

Despite the significant potential of doping with multiple ions in biomedical applications, there is still limited research in this area, where most of these studies have been focused on structural modifications applied to bioceramics. Consequently, this field of research holds a promising future with numerous opportunities, and it is expected that further exploration and investigation will be conducted, particularly focusing on the biological attributes of multi-doped bioceramics.

## 5. Potential co-dopants for biomedical silicates

Even though the various combinations of co-doping have been introduced to regulate the biological characteristics of silicate bioceramics, there is still ample room for further investigation and improvement in this field. The selection of dopants should be based on the existing knowledge, but it should also be considered that certain cases have shown adverse or unexpected effects compared to their mono-dopants. For example, diopside co-doped with Sr and F exhibited a lower ability to deposit apatite compared to the samples with mono-doping or no doping [117]. Therefore, it is crucial to explore different dopant systems and interactions. Figure 8 provides a summary of the dopants previously used in silicate bioceramics, either individually or in combination, and their predominant impacts on the bio-properties of silicates. According to this database, several potential co-dopants for biomedical silicate systems are suggested in this section for future research efforts, with the aim of further improving bio-performance, as summarized in Table 3. Some of these co-dopants have previously demonstrated successful results in other bioceramics.

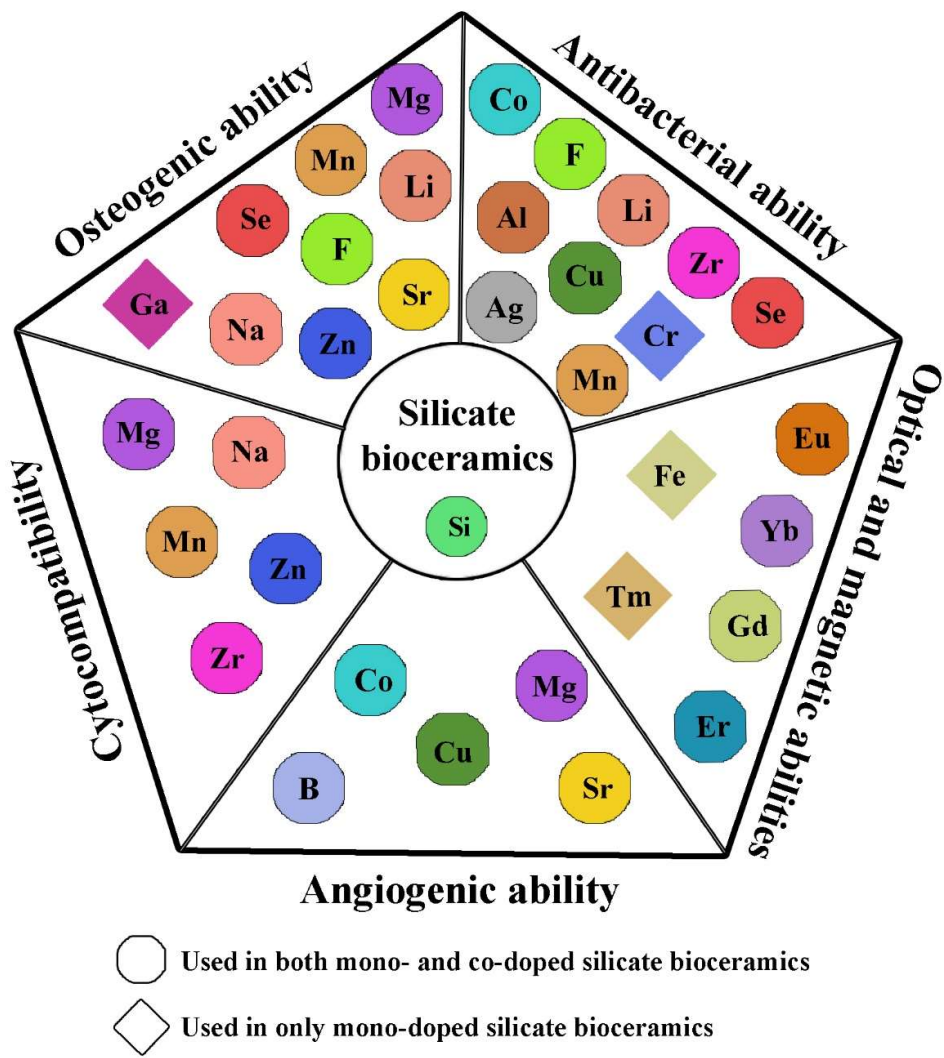

Figure 8. A summary of the dopants used in silicate bioceramics under mono- and co-doped conditions.

Table 3. A review on possible co-dopants and their expected modifications on the properties of silicate bioceramics

| Possible co-dopants in silicates | Expected modifications | Evidences | Reference for evidences |
|---|---|---|---|
|  |  |  |  |
| **Cr-Zn** | Antibacterial effect, bioactivity, cell activity, and osteogenesis ability | Ability to form an apatite layer on HA | [227] |
| **Co-F** | Angiogenesis and Osteogenesis improvements | Improved osteogenesis ability on HA | [228] |
| **Co-Mg** | Angiogenesis and cell functionality | Enhanced MG-63 cell viability and higher VEGF expression on HA | [229] |
| **Co-Zn** | Angiogenesis, antibacterial, and osteogenesis abilities | - | - |
| **Cu-Mg** | Angiogenesis, antibacterial, and osteogenesis abilities | Enhancement of antibacterial effect, cell proliferation, and mechanical properties on HA | [230] |
| **F-Zn** | Antibacterial effect and cell activity | - | - |
| **Fe-X** | Magnetic properties, degradability, and cell activity | Targeting cancerous cells and impact on magnetic behaviors of Ag-Fe doped HA nanoparticles | [231] |
| | | Improved cell functionality of Fe-Sr co-doped BCP | [232] |
| | | Blood compatibility, bioactivity, and antimicrobial effect of Fe-Zn co-doped nanocrystalline HA | [233] |
| **Li-Sr** | Osteogenesis ability | - | - |
| **Mn-Sr** | Osteogenesis ability and cell functionality | Improved cell activity toward MC3T3-E1 on HA films | [234] |

| | | Higher MG63 cell proliferation and differentiation on brushite | [235] |
|---|---|---|---|

## 5.1. Chromium-Zinc

Serin et al. [236] demonstrated that doping of Cr into bioglass can impart antibacterial properties, improve cell activity, and enhance bioactivity. However, it should be noted that high amounts of Cr can be cytotoxic. Furthermore, Zn has antibacterial abilities and can promote osteogenesis [38]. This pair has previously been doped into HA by Tautkus et al. [227], illustrating the formation of apatite in Cr-Zn co-doped HA. The co-incorporation of Cr and Zn into silicate bioceramics has not yet been addressed, and it holds the potential to enhance bone formation in these materials.

## 5.2. Cobalt-Fluorine

Efficient bone regeneration requires both angiogenesis and osteogenesis capabilities. Therefore, a promising approach is to consider dopants that can address these properties. Co is well-known for promoting angiogenesis in bioceramics, while F has the ability to induce osteogenesis. However, Co can have adverse effects on osteogenesis and reduce cell viability at high concentrations. To counterbalance these negative effects, F can be utilized. Birgani et al. [228] assessed the impact of Co-F co-doping on apatite bioceramics. Their findings indicated that mono-doped Co had an adverse effect on ALP activity, resulting in lower osteogenic properties compared to the non-doped samples. However, the co-doped apatite demonstrated superior osteogenesis compared to the non-doped samples, indicating the ability of F to offset the adverse effect of Co on osteogenesis. Additionally, the co-doped specimens exhibited improved angiogenic behavior compared to the non- and mono-doped specimens. It highlights the potential of Co-F co-doping in silicate bioceramics, an approach that has not been reported yet.

### 5.3. Cobalt-Magnesium

The combination of Co-Mg offers another set of co-dopants that can confer both angiogenesis and osteogenesis properties. Co has an outstanding ability to promote angiogenesis, while Mg plays a vital role in cell functionality and can aid in the vascularization process. Thus, the combination of Co and Mg can dramatically affect the angiogenesis properties of bioceramics. Kulanthaivel et al. [229] fabricated Co-Mg co-doped HA and observed enhanced MG-63 cell viability in the co-doped specimens. Additionally, the co-doped specimens exhibited a higher expression of VEGF, a key factor in the inducement of angiogenesis.

### 5.4. Cobalt-Zinc

Co has the capability to promote angiogenesis, and Zn promotes the osteogenesis ability of bioceramics, while both also have antibacterial effects. Therefore, the co-incorporation of Co-Zn into bioceramics appears to be a suitable option for bone regeneration. Nevertheless, to our knowledge, there are no reports on co-doping of silicate bioceramics with Co and Zn. Considering that angiogenesis is a crucial factor for improving osteogenesis, it is plausible that co-doping of silicate bioceramics with Co and Zn would enhance the osteogenesis ability compared to the non-doped or mono-doped silicates. Additionally, Co-Zn co-doping can induce synergistic antibacterial effects, while Zn may offset the adverse effect of Co on cell viability and osteogenesis.

### 5.5. Copper-Magnesium

The effect of Mg on various types of silicate bioceramics has been investigated either individually or in combination with other elements, demonstrating its ability to promote bone regeneration. However, magnesium has limited effects on vascularization and bactericidal properties. On the other hand, Cu is known for its angiogenic and antibacterial activities. Despite the favorable effects of both Cu and Mn, the Cu-Mn combination remains an intriguing option for doping. Veljovic et al. [230] investigated the addition of this pair to HA, suggesting that it can enhance antibacterial properties, cell proliferation, and mechanical characteristics. Nevertheless, to our knowledge, no research has been reported on utilizing this pair as co-dopants in silicate bioceramics

### 5.6. Fluorine-Zinc

F and Zn both have shown the ability to alter the antibacterial and cell activities of bioceramics. Therefore, the co-incorporation of F and Zn can further enhance these attributes in bioceramics compared to the non-doped or mono-doped samples. However, to our knowledge, there is currently no research on the impacts of F-Zn co-doping in silicate bioceramics. As evidence, Uysal et al. [237] demonstrated that co-doped HA exhibits improved mechanical properties (microhardness and fracture toughness) and cell proliferation.

### 5.7. Iron-X

The addition of Fe as a dopant to bioceramics can alter their magnetic properties, degradability, and cell activity, as shown in studies on akermanite [238] and hardystonite [239, 240]. Therefore, Fe-doped silicate bioceramics are suitable candidates for hyperthermia and tumor therapy. Despite the significant potential of Fe and researches conducted on Fe mono-doped silicate bioceramics, the application of Fe as a co-doping element in silicate bioceramics has not been investigated. Veerla et al. [231] demonstrated that Ag-Fe co-doped HA nanoparticles can target cancerous cells, while being safe for normal cells. Furthermore, the addition of Fe dramatically impacted the magnetic behaviors of the nanoparticles. Basu et al. [232] showcased the improved cell functionality of Fe-Sr co-doped BCPs toward epithelial cells, and Jayapalan et al. [233] indicated the blood compatibility, bioactivity, and antimicrobial abilities of Fe-Zn co-doped nanocrystalline HA.

### 5.8. Lithium-Strontium

Li and Sr both have been shown to significantly induce bone formation. One common mechanism by which these dopants promote new bone formation is through cell signaling. Cell signaling is effective to enhance the bone formation ability of silicate bioceramics, and researchers have investigated the use of mono-doping for this purpose. Li affects different cellular signaling pathways, such as the Wnt/β-catenin pathway, critical for bone regenerative applications [241]. Additionally, Sr can enhance osteogenesis by activating the AKT/mTOR signaling pathway [242] and Wnt/β-catenin pathway [243]. No reports have been published on the co-substitution of bioceramics with Li and Sr, to our knowledge. However, considering the potential synergistic and additive effects of this pair on cell signaling, as well as the other beneficial properties that each dopant brings to the bioceramic, it presents an exciting prospect for bone regenerative applications.

### 5.9. Manganese-Strontium

Mn plays a vital role in bone evolution via stimulating the expression of osteogenic markers such as ALP. Sr is known for its stimulatory impact on bone regeneration. While Sr has been extensively studied, the application of Mn as a dopant is limited. It is plausible that a combination of these elements can have a synergistic effect on bone regeneration, given their individual abilities to induce bone formation. However, to our knowledge, there is no work reported on Mn-Sr co-doped silicate-based bioceramics. Nonetheless, Huang et al. [234] indicated that Mn-Sr co-doped HA films on titanium surfaces exhibited superior cell activity toward MC3T3-E1 cells compared to pure HA coatings. Additionally, Mn-Sr co-doped brushite cements fabricated by Marote et al. [235] revealed higher MG63 cell proliferation and differentiation compared to non-doped or Sr mono-doped cements.

## 6. Technological landscape of co-doped silicate bioceramics

Bioceramics have long been integrated into the biomedical industry, particularly in the realm of therapeutic hard tissue applications. For instance, calcium phosphates are widely used in pharmaceutical and bone-related applications, while alumina and zirconia serve as primary components of hip and femur prostheses. Several doped bioceramics have also been commercialized, including Si-containing calcium alkali orthophosphate (Osseolive®) and hydroxyapatite (Actifuse®). Additionally, there is a multitude of silicate bioceramics on the market, particularly glasses such as BioMin®C and OssiMend®, as reviewed by Shearer et al. [244]. Despite the promising results obtained from available scientific studies, the application of co-doped silicate bioceramics has not yet reached the commercialization level, to our knowledge. This can be due to various reasons, including: 1) the complexity of interactions between/among dopants with each other, with the ceramic matrix, and even with the production process and 2) the lack of sufficient *in vivo* and clinical knowledge on this issue. Nevertheless, we believe that the commercialization of these products is not far off, as there are already some granted patents on co-doped silicate bioceramics, with a focus on Sr as one of the two elements doped. The pairs that have paved their way to patents are B-Na [35], Zn-Mg and B-Mg [103], and Eu-Sr, Ag-Sr, Zn-Sr, and Cu-Sr [114], which are discussed in detail in Section 3 under their relevant subsections. It suggests that further research and development in this area could lead to the introduction of new and effective biomedical products in the near future.

## 7. Conclusions and future directions

This article presented a comprehensive review on the effect of co-doping on the bio-performance of silicate bioceramics. More specifically, it provided a summary of the biological impacts of elements used as dopants in co-doped silicates (including Al, B, Ce, Co, Cu, Er, Eu, F, Gd, Li, Mg, Mn, , Se, Ag, Na, Sr, Tm, Yb, Zn, and Zr), as well as co-dopants incorporated in silicates (including Al-Mg, Al-Sr, B-Co, B-Cu, B-Mg, B-Na, Ce-Ag, Co-Li, Co-Sr, Cu-Ag, Cu-Sr, Er-Yb, Eu-Gd, Eu-Sr, F-Sr, Li-Ag, Mg-Ag, Mg-Sr, Mg-Zn, Mn-Zn, Se-Sr, Ag-Sr, Ag-Zn, Ag-Zr, Sr-Zn, Zn-Zr, and Mg-Sr-Zn). It explored the summative, synergistic, or antagonistic interaction of the elements with the host ceramic and each other, providing suggestions for future research directions (such as Cr-Zn, Co-F, Co-Mg, Co-Zn, Cu-Mg, F-Zn, Fe-X, Li-Sr, and Mn-Sr). From a comparative viewpoint, Sr likely emerges as the most dominant dopant for enhancing osteogenesis and inhibiting osteoclast activity. Ag and Cu appear to be the optimal choices for the antibacterial effect, considering cytotoxicity thresholds that should be met for both. Mg seems to be the most notable dopant for enhancing cell activity, while B and Cu can significantly promote angiogenesis without the side effects of dopants such as Co. However, some dopants have been shown to be less effective than anticipated. For example, while Zn promotes cellular and antibacterial activities, it significantly retards apatite and bone formation. It was concluded that the incorporation of co-dopants into silicates often provides the summative or synergistic effects of dopants, enhancing properties compared to mono-doped counterparts. However, it is not a universal rule; for example, F-Sr co-doping resulted in inferior properties compared to Sr and F mono-doped samples in terms of an antagonistic interaction. Therefore, to design new co-doping systems, in addition to a comprehensive understanding of mono-doping effects, theoretical, computational, and/or experimental assessments should be conducted on co-doped samples to fully comprehend interactions between/among different dopants and achieve optimal formulations. Apart from co-dopant species, consequences also depend on the composition, crystallinity, and specific surface area of the host material, as these factors significantly affect the release rate of incorporated ions. These complexities as well as the lack of sufficient *in vivo* studies on these systems have slowed down their clinical and technological translations. Future research on co-doping of silicate bioceramics is expected to focus in utilizing multiple co-dopants to further improve their behavior. The integration of co-doping and other therapeutic intervention agents, such as drugs or growth factors, can be considered to develop more effective consequences. Additionally, the presence of several granted patents in this area and the availability of mono-doped commercial products suggest a promising landscape for the clinical commercialization of these bioceramics.